\documentclass[journal,draftcls,onecolumn,12pt,twoside]{IEEEtran}
\usepackage[hidelinks]{hyperref}
\usepackage[utf8]{inputenc}
\usepackage[T1]{fontenc}
\usepackage{placeins}
\usepackage{microtype}
\usepackage[dvips]{graphicx}
\usepackage{caption}
\graphicspath{ {images/} }
\usepackage{xcolor}
\usepackage{times}
\usepackage{amsmath}
\usepackage{float}
\usepackage{epstopdf}
\usepackage{amssymb}
\usepackage{blindtext}
\usepackage{enumitem}
\usepackage{xcolor}
\usepackage{amsmath}
\usepackage{epstopdf}
\usepackage{array}
\usepackage{multirow}
\usepackage{algorithm}
\usepackage{cite}
\usepackage{tikz}
\usetikzlibrary{shapes}
\usetikzlibrary{plotmarks}
\usetikzlibrary{spy}
\usepackage{import}
\usepackage{pgfplots}
\usepackage{mathtools}
\usepackage{multirow} 
\usepackage{adjustbox}
\usetikzlibrary{spy}
\usepackage[utf8]{inputenc}
\usepackage{tikz}
\usetikzlibrary{matrix,decorations.pathreplacing}
\usetikzlibrary{shapes}
\usepackage[utf8]{inputenc}
\usepackage{amsmath}
\usepackage{amssymb}
\usepackage{amsthm}
\usetikzlibrary{plotmarks}
\usepackage{pgfplots}
\usetikzlibrary{spy}
 \usepackage{pstricks-add}
 \usepackage{epsfig}
 \usepackage{pst-grad} 
 \usepackage{pst-plot} 
 \usepackage[space]{grffile} 
 \usepackage{etoolbox}
 	\definecolor{darkblue}{rgb}{0.0, 0.0, 0.55}
 \usepackage{tikz}
 \usepackage{pgfplots}
\usepackage[a4paper,margin=2cm]{geometry}
\usepackage{forest}
\usetikzlibrary{trees,positioning,shapes,shadows,arrows.meta}

\usepackage[labelformat=simple]{subcaption}

\usepgfplotslibrary{external}
\usepackage{algpseudocode}

\usepackage{etoolbox}
\AtBeginEnvironment{algorithm}{\algoequations}
\AtEndEnvironment{algorithm}{\restoreequations}
\newcounter{algosavedequation}
\newcommand{\algoequations}{%
  \setcounter{algosavedequation}{\value{equation}}%
  \setcounter{equation}{0}%
  \renewcommand{\theequation}{A.\arabic{equation}}%
}
\newcommand{\innerproduct}[2]{\langle #1, #2 \rangle}
\newcommand{\restoreequations}{%
  \setcounter{equation}{\value{algosavedequation}}%
}
\DeclareMathOperator*{\argmax}{argmax}
\DeclareMathOperator*{\argmin}{argmin}
\makeindex 
\ifCLASSINFOpdf
  
\else

\fi

\definecolor{mygr}{HTML}{e6e6e6}
\definecolor{bblue}{HTML}{4F81BD}
\definecolor{rred}{HTML}{C0504D}
\definecolor{ggreen}{HTML}{9BBB59}
\definecolor{ppurple}{HTML}{9F4C7C}
\newtheorem{ex}{Example}

\begin{document}

\tikzset{new spy style/.style={spy scope={%
			magnification=2.8,
			size=1.25cm,
			connect spies,
			every spy on node/.style={
				rectangle,
				draw,
			},
			every spy in node/.style={
				draw,
				rectangle,
			}
		}
	}
}

\newcolumntype{L}[1]{>{\raggedright\let\newline\\\arraybackslash\hspace{0pt}}m{#1}}
\newcolumntype{C}[1]{>{\centering\let\newline\\\arraybackslash\hspace{0pt}}m{#1}}
\newcolumntype{R}[1]{>{\raggedleft\let\newline\\\arraybackslash\hspace{0pt}}m{#1}}
\pgfdeclarelayer{background}
\pgfdeclarelayer{foreground}
\pgfsetlayers{background,main,foreground}
\newcommand{\oeq}{\mathrel{\text{\sqbox{$=$}}}}
\setlength{\textfloatsep}{0.1cm}
\setlength{\floatsep}{0.1cm}
\tikzstyle{int}=[draw, fill=white!20, minimum size=2em]
\tikzstyle{init} = [pin edge={to-,thin,black}]

\title{ADMM-based Detector for Large-scale MIMO Code-domain NOMA Systems }
\author{  \IEEEauthorblockN{\small{Vinjamoori Vikas$^{1}$, Kuntal Deka$^{1}$,  Sanjeev Sharma$^{2}$, and  A. Rajesh$^{1}$}} \\
$^1$Indian Institute of Technology Guwahati, India,   $^2$Indian Institute of Technology (BHU) Varanasi, India
}

\maketitle  

\begin{abstract}
Large-scale multi-input multi-output (MIMO) code domain non-orthogonal multiple access (CD-NOMA) techniques are one of the potential candidates to address the next-generation wireless needs such as  massive connectivity, and high reliability. This work focuses on two primary CD-NOMA techniques: sparse-code multiple access (SCMA) and dense-code multiple access (DCMA). One of the primary challenges in implementing MIMO-CD-NOMA systems is designing the optimal detector with affordable computation cost and complexity.   This paper proposes an iterative linear detector based on the alternating direction method of multipliers (ADMM). First, the maximum likelihood (ML) detection problem is converted into a sharing optimization problem. The set constraint in the ML detection problem is relaxed into the box constraint sharing problem. An alternative variable is introduced via the penalty term, which compensates for the loss incurred by the constraint relaxation. The  system models, i.e., the relation between the input signal and the received signal, are reformulated so that the proposed sharing optimization problem can be readily applied.
  The  ADMM is a robust algorithm to solve the sharing problem in a distributed manner. The proposed detector leverages the distributive nature to reduce per-iteration cost and time. An ADMM-based linear detector is designed for three  MIMO-CD-NOMA systems: single input multi output CD-NOMA (SIMO-CD-NOMA), spatial multiplexing CD-NOMA (SMX-CD-NOMA), and spatial modulated CD-NOMA  (SM-CD-NOMA). The impact of various system parameters and ADMM parameters on computational complexity and symbol error rate (SER) has been thoroughly examined through extensive Monte Carlo simulations.  

\end{abstract}

\IEEEpeerreviewmaketitle

\section{Introduction}
\subsection{Motivation}
The existing literature extensively studies two types of code domain non-orthogonal multiple access (CD-NOMA) systems.
The first type is sparse coded CD-NOMA systems, which utilize low-density spreading sequence design (e.g., low-density signatures (LDS)) or sparse codebook design (e.g., sparse code multiple access (SCMA)) \cite{taherzadeh2014scma}.
The second type is densely coded CD-NOMA systems, which utilize dense spreading sequence design (e.g., overloaded code-division multiple access (CDMA)) or dense codebook design (e.g., dense code multiple access (DCMA)) \cite{Liu}. Throughout the paper, dense codebook-based NOMA is referred to as DCMA, and dense spreading sequence-based NOMA is referred to as overloaded CDMA.


SCMA and DCMA are critical enabling technologies to improve spectral efficiency and massive connectivity by overloading multiple user equipment's (UEs) data on a single resource element (RE). These techniques exploit the codebooks' multi-dimensional constellation (MDC) coding gain. Spreading sequence-based NOMA (LDS, overloaded CDMA) does not benefit from the coding gain of the MDC. Thus,  codebook-based NOMA techniques offer substantial error rate performance gains over the spreading sequence-based NOMA. 

Multi-input multi-output (MIMO) technology is another crucial enabler in improving spectral efficiency and system reliability \cite{chockalingam}. CD-NOMA systems using a  MIMO system (MIMO-CD-NOMA) offer even greater spectral efficiency, reliability, and massive connectivity improvements.
This paper considers three types of uplink (UL) MIMO-CD-NOMA systems: (1) Single-input multi-output (SIMO) aided CD-NOMA, where each UE has a single antenna \cite{Lim}; (2)~ Spatial multiplexing-CD-NOMA (SMX-CD-NOMA), where each UE is equipped with multiple transmit antennas \cite{Elkawafi}; and (3) Spatial modulated CD-NOMA (SM-CD-NOMA), where each UE activates a single transmit antenna out of multiple transmit antennas at the UE \cite{Pan}.
The primary challenge in implementing MIMO-CD-NOMA systems lies in designing a multiuser signal detector that offers excellent performance while maintaining affordable complexity.



  The  message passing algorithm (MPA)  is usually used in SCMA detection \cite{Zhang, Yang_Lin }. However, the MPA for MIMO-SCMA signal detection \cite{Dai, Du} needs to improve its computational complexity. MPA detection over the MIMO-SCMA system uses an extended factor graph due to the additional antennas at the  UEs and the BS \cite{KD_MIMO}. The MPA exhibits exponential complexity with the codebook size ($M$), the number of UEs ($J$), and the number of transmit antennas. Thus, MPA becomes impractical for highly overloaded  SCMA systems over large-scale MIMO systems.
Further, MPA often faces convergence issues when the factor graph contains cycles. To overcome the limited diversity issues (due to sparsity) in SCMA, researchers have recently started designing the DCMA systems \cite{Liu}. However, MPA demands sparsity for the accurate detection of codewords. DCMA codewords do not hold sparsity properties. Due to enormous short cycles in the DCMA factor graph, MPA is no longer a suitable detector for DCMA systems. Thus, an alternate detection algorithm must be developed to achieve the full diversity offered by DCMA with minimal detection complexity.
  
    The generalized sphere decoder (GSD) is used as a detector for a spread sequence-based DCMA system by considering a single antenna at both the base station (BS) and the UE \cite{Liu}. The GSD works on principles of the tree search algorithm. Its complexity depends on the number of tree nodes and floating point operations (FLOPs) at each node. The number of tree nodes will increase multi-fold in the codebook-based DCMA systems. 
  Moreover, the complexity of GSD increases rapidly for large-scale MIMO DCMA systems. So, it is not an efficient detector for large-scale MIMO CD-NOMA systems. Designing a low-complexity multiuser detector (MUD) for large-scale MIMO-aided CD-NOMA systems with higher overloading factors ($\lambda$) and modulation order/codebook-size ($M$)  is becoming essential.
  The existing research studies mainly nonlinear detectors (MPA and sphere decoder) for CD-NOMA systems \cite{Zhang, Wei}. On the other hand, minimum mean square error (MMSE) and zeros forcing (ZF) are commonly used linear detectors in standalone  MIMO systems\cite{Yang}. These detectors are simple to implement and they exhibit low computational complexity over nonlinear detectors. However, these detectors show poor symbol error rate (SER) performance over CD-NOMA systems with high $\lambda$ values. In addition, they fail to detect the CD-NOMA signals as  $M$ increases. Thus, designing an efficient detector with low complexity is mandatory to fully unfurl the benefits of MIMO CD-NOMA systems with a practically viable approach.
  Further, the detector must perform better than the  existing linear detectors for large-scale CD-NOMA parameters ($\lambda$, $M$). 
  \vspace{-1mm}
\subsection{Related prior works}
   Recently, the ADMM has been widely used to solve convex and non-convex problems in a distributed manner\cite{boyd}. Glowinski and Marrocco first proposed the ADMM in the mid-1970s. Later on, Boyd \textit{et al}. rigorously discussed various complex optimization problems that can be solved using ADMM in their tutorial \cite{boyd}. The ADMM is formed by the composition of dual ascent and the method of multipliers. The dual decomposition properties of dual ascent make the ADMM solve in a distributed manner. ADMM has superior convergence properties and often is able to converge without the requirement of strict convexity. 
  A comprehensive treatment of ADMM and its advancements are detailed in ~\cite{Zhouchen}.

  The ADMM is extensively applied to solve the linear programming (LP) problem in low-density parity-check code's (LDPC) decoding\cite{Barman, Banihashemi}. The authors here have also shown that the ADMM significantly reduces the complexity of the decoder compared to the belief propagation (BP) methods. 
    The MUD problem of UL grant-free NOMA is solved by using ADMM \cite{Cirik}. The ADMM-based infinity norm (ADMIN)  iterative linear detector is proposed for massive MIMO systems\cite{Shahabuddin}. The authors have also proposed VLSI architecture for the ADMIN detector. The results show that ADMIN outperforms all the linear detectors in terms of SER performance and low hardware cost compared to the nonlinear BP-based detectors. Similarly, an ADMM-based QAM signal detector for massive MIMO systems is proposed  \cite{Zhang_Quan}. A distributed penalty sharing (DPS) ADMM method is designed to convert the maximum-likelihood (ML) detection problem into sharing optimization problem. This method shows good performance and complexity trade-offs for massive MIMO systems. The authors in \cite{Zhang_Quan, Shahabuddin} show that the ADMM-based detector performs MMSE equalization in the first iteration. The performance of the ADMM improves over MMSE as iterations progress.
 
 \vspace{-2mm}
 \subsection{Contributions:}
 This paper proposes an ADMM-based iterative linear detector that solves the MUD problem of large-scale MIMO-aided CD-NOMA systems.  It exhibits the best trade-off between the SER performance and computational complexity for SCMA and DCMA systems.    We have formulated the highly computationally complex ML detection problem of large-scale MIMO-CD-NOMA systems as the non-convex distributed optimization problem. To the best of the authors' knowledge, this is the first attempt to convert the ML detection problem of CD-NOMA into a sharing problem. Further, the  ADMM approach is applied to solve the proposed sharing problem using distributed optimization. The proposed ADMM-based detector is guaranteed to improve the error rate performance over linear detectors with lower computational complexity than nonlinear detectors. The main contributions of this work are listed below:\\
  \vspace{-5mm}
 \begin{itemize}
   \item The ML MUD problem is nondeterministic polynomial-time hard (NP-hard) and can be solved using exhaustive search \cite{Wing-Kin}. However, the complexity of the exhaustive search increases exponentially as the number of UEs and codebook size increase.
The ML problem is transformed into a non-convex sharing optimization problem to overcome this challenge. An efficient distributed optimization method is then applied to solve the proposed sharing problem.
\item The CD-NOMA system models, i.e., the relationships between the input signal and received signal, are reformulated so that the sharing ADMM algorithm can be readily applied for detection. For SIMO-CD-NOMA, the signal received by multiple antennas at BS is considered as a single observation vector for detection. For SMX-CD-NOMA and SM-CD-NOMA, resource-wise processing of the received signal is carried out during ADMM detection.
\item A  low-complexity iterative linear detector is designed via the ADMM approach to solve the sharing optimization problem for large-scale MIMO-CD-NOMA systems.  ADMM solves the sharing optimization problem through a distribution optimization framework.
Indeed, distributive nature of sharing ADMM allows parallel processing in multiuser detection, reducing per-iteration computational time.
\item  The proposed ADMM-based detector is applied to large-scale MIMO (SIMO, SMX-MIMO, and SM-MIMO) aided   CD-NOMA systems. Thus, the proposed single  ADMM-based MUD is capable of solving  the detection problem of various CD-NOMA systems. 
\item The complexity of the proposed ADMM-based detector is independent of the CD-NOMA system's codebook size/modulation order ($M$), unlike the conventional MPA detector. Thus, it can be applied to large-size codebooks. The impact of the different MIMO-CD-NOMA system parameters is also thoroughly examined. The ADMM-based detector exhibits polynomial complexity with all other parameters, such as the numbers of antennas, UEs, and REs.
  
\item  The comparison of the error rate performance and receiver complexity of the proposed ADMM-based detector with all other conventional detectors is thoroughly examined. Comprehensive Monte Carlo simulations indicate that the ADMM-based detector substantially reduces the computational complexity  while maintaining an error performance  comparable to the MPA in SCMA detection. For spread sequence-based DCMA, the ADMM gives superior performance than GSD.
 \end{itemize}
 \vspace{-5mm}

 \subsection{Organization:}
 The paper is  organized  as follows.  The preliminaries of the existing CD-NOMA system models and sharing ADMM problem are discussed in Section \ref{sec2}.  The proposed  system models and ADMM-based detection  for UL MIMO-CD-NOMA systems are provided in Section \ref{sec.2}. Section ~\ref{sec4} presents  a detailed  computational complexity analysis. The simulation results for different MIMO-CD-NOMA systems are discussed in Section \ref{sec.5}. Section \ref{sec6} concludes the paper and mentions future scopes.

\noindent \textit{Notations:} Lower case, bold lower case, and bold upper case letters denote scalars, vectors, and matrices, respectively. $(\cdot)^T$ and $(\cdot)^H$ denote transpose and Hermitian transpose, respectively.  $\Vert (\cdot) \Vert$ denotes the Euclidean norm of a vector. $\prod_{[-\alpha,\alpha]} (\cdot)$ denotes the Euclidean projection onto the interval $[-\alpha,\alpha]$. $\innerproduct{\cdot}{\cdot}$ denotes the inner product, and ${\text{Re}}(\cdot)$ denotes the real part of a complex variable.  $\mathbb{R}^n$ and $\mathbb{C}^n$ denote $n$-dimensional real and complex vector spaces, respectively. $\mathcal{CN}(0,\sigma^2)$ denotes complex Gaussian distribution with zero mean and variance $\sigma^2$.

\section{Preliminaries}\label{sec2}

This section mainly discusses the existing CD-NOMA system models and steps to solve the sharing optimization problem via the ADMM approach. The CD-NOMA techniques can be broadly divided into sparse-coded and dense-coded NOMA. Low-density signature (LDS) \cite{Hoshyar}  and SCMA \cite{Nikopour} belong to the first group, whereas   overloaded CDMA \cite{Liu} and  DCMA  \cite{Liu} belong to the second group. 
LDS  and overloaded CDMA are sequence-based CD-NOMA techniques where mapping and spreading are performed separately. LDS utilizes sparse sequences, while overloaded CDMA employs dense sequences. 
 These techniques suffer from error performance loss due to limited coding gain. 
 On the other hand, SCMA  and DCMA are codebook-based CD-NOMA techniques. In these systems, the data of each UE is mapped to a multi-dimensional codeword. SCMA utilizes sparse codewords, while DCMA applies dense codewords. These techniques offer error performance benefits through the use of MDCs \cite {Boutros}.
 
 
  

\vspace{-5mm}
\subsection{SCMA system model}
\begin{figure*}[h]
\centering
\begin{subfigure}[b]{0.31\textwidth}
\centering
\includegraphics[width=\textwidth]{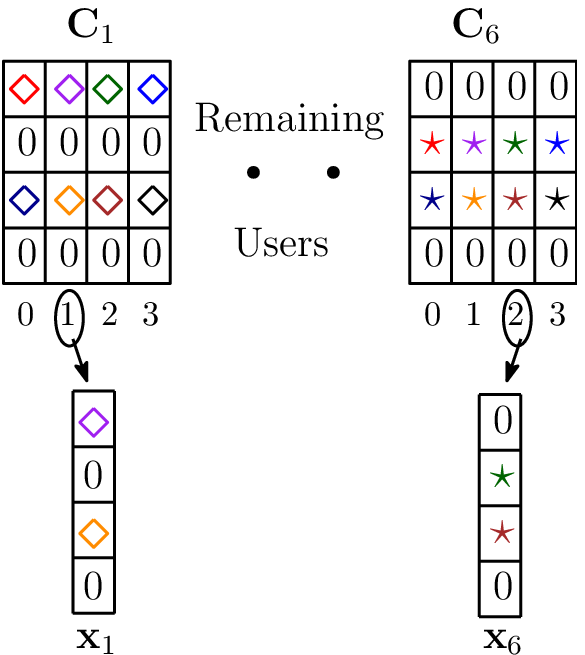}      
\caption{\footnotesize{SCMA codebook structure }}
\label{scma_model}
\end{subfigure}\hspace{2em}  \vline   \hspace{2em}
\begin{subfigure}[b]{0.32\textwidth}
\centering
\includegraphics[width=\textwidth]{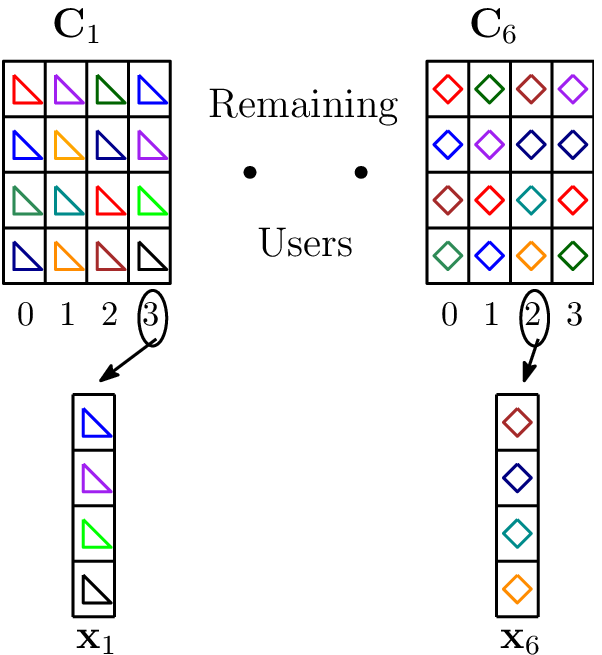}      
\caption{\footnotesize{DCMA codebook structure}}
\label{dcma_model}
\end{subfigure}
\caption{CD-NOMA codebook structure for $J=6$ and  $K=4$.}
\end{figure*}

SCMA is a sparse codebook-based NOMA technique. Each UE's data is sent in the form of sparse codewords.
Consider a CD-NOMA system having $J$  UEs accessing $K$  REs, where $J>K$ ensures the overloading nature of SCMA. Each UE in the SCMA system has  access to $d_{\mathrm v}<K$  active REs among $K$ REs. Thus, $d_{\mathrm v}$ non-zero elements are present in the $K$-dimensional sparse codewords. The SCMA encoder maps each UE bitstream to a sparse codeword $\mathbf{x}^{K\times 1}$ in a pre-designed SCMA codebook $\mathcal{X}^{K\times M}$ of the respective UE. The sparse nature of SCMA codewords enforces $d_{\mathrm f}< J$ number of overlapping UEs on each RE.  The diversity order of SCMA is essentially limited to $d_{\mathrm v}$, i.e., far less than the maximum possible diversity order $K$. For example, the  codebook structure for  150~\% overloaded SCMA system ($d_{\mathrm v}=2$, and $d_{\mathrm f}=3$) is shown in Fig. \ref{scma_model}. The received signal vector is given by
\vspace{-0.15in}
\begin{align}\label{scma.eq1}
    \mathbf{r}= \sum_{j=1}^J \text{diag}(\mathbf{h}_j)\mathbf{x}_j+\mathbf{w},
\end{align}
where 
\begin{itemize}
	\item $\mathbf{w}\sim \mathcal{CN}(0,\sigma^2\textbf{I}_K)$ denotes the additive white Gaussian noise (AWGN) at the receiver.
	\item $\mathbf{h}_j\sim \mathcal{CN}(0,\textbf{I}_K)$ denotes the Rayleigh fading  channel vector for the $j$th UE.
	\item  $\text{diag}({\mathbf{h}}=\left[h[1],\ldots, h[k], \ldots, h[K]\right]^T)$ denotes diagonal matrix and  $h[k]$ is the $k$th diagonal element.
 \item $\mathbf{x}_j$ is the $K$-dimensional codeword of $j$th UE.
\end{itemize}

\vspace{-7mm}

\subsection{DCMA system model}
The error performance of SCMA suffers from limited diversity order.
This limitation of SCMA can be overcome by using dense codebooks in the DCMA system \cite{Liu}. Each UE in the DCMA system has access to all $K$  REs. No zero entries are present in the $K$-dimensional dense codewords.   Therefore, the DCMA system exploits the full diversity of multi-dimensional codewords.  For example,  the codebook structure of 150~\% overloaded DCMA system   is shown in Fig. \ref{dcma_model}. The received signal vector for the DCMA system is similar to that of the SCMA system, as described in  (\ref{scma.eq1}).
In addition to the  codebook-based DCMA system, the spread sequence-based DCMA  can be designed with non-orthogonal spreading sequences. The idea is similar to the overloaded CDMA systems. The number of UEs is larger than the length of the spreading sequence. Uni-modular spreading sequences have been used in this paper to achieve full diversity of the dense sequences \cite{Liu}.  
\vspace{-2mm}
\subsection{Shared ADMM problem}\label{sec.II(C)}
This subsection introduces the fundamental concepts of solving sharing problems using ADMM within a distributed optimization framework. A novel method, referred to as ADMM, was proposed to address both convex and non-convex sharing optimization problems \cite{boyd}. The subsequent subsection delves into the details and ideas behind the sharing ADMM problem.
The ADMM is formed by combining the superior properties of dual ascent and the method of multipliers. This combination ensures the robustness of ADMM \cite{boyd}. 
The generic sharing  problem  is
\begin{align}
\min\quad \sum_{i=1}^N f_i(\mathbf{x}_i)+g\left(\sum_{i=1}^N \mathbf{x}_i\right),\label{eq.ADMM1}
\end{align}
where $\mathbf{x}_i\in \mathbb{R}^n, i=1,\hdots, N$, the associated local cost function $f_i(\mathbf{x}_i) (f_i: \mathbb{R}^n\rightarrow \mathbb{R})$ of subsystem $i$ is handled by processor $i$, and $g$ $(g: \mathbb{R}^n\rightarrow \mathbb{R})$ is the shared objective, whose  argument is the sum of $N$ variables.  Each variable $\mathbf{x}_i$ is involved in minimizing the individual cost $f_i(\mathbf{x}_i)$, as well as the shared objective $g\left(\sum_{i=1}^N \mathbf{x}_i\right)$. The sharing problem can be converted into an ADMM problem by introducing an alternative variable, $\mathbf{z}_i\in \mathbb{R}^n$. Thus, the cost function is minimized over $\mathbf{x}_i$ and $\mathbf{z}_i$, alternatively. The problem (\ref{eq.ADMM1}) is converted to the following problem:
\begin{align}\label{eq.ADMM2}
\min\quad &\sum_{i=1}^N f_i(\mathbf{x}_i)+ g\left(\sum_{i=1}^N \mathbf{z}_i \right)\\
\textrm{s.t.}\quad & \mathbf{x}_i-\mathbf{z}_i=0\quad i=1,\hdots,N.\nonumber 
\end{align}
 The augmented Lagrangian function for (\ref{eq.ADMM2}) can be written as 
\begin{align}\label{eq.ADMM-aug}
\mathcal{L}\left(\{\mathbf{x}_i,\mathbf{z}_i,\mathbf{y}_i\}_{i=1}^N\right)=\sum_{i=1}^N f_i(\mathbf{x}_i)+g\left(\sum_{i=1}^N \mathbf{z}_i \right) +\sum_{i=1}^N \innerproduct{\mathbf{x}_i-\mathbf{z}_i}{\mathbf{y}_i}+\frac{\rho}{2}\sum_{i=1}^N \Vert \mathbf{x}_i-\mathbf{z}_i\Vert_2^2
\end{align}
where $\mathbf{x}_i, \mathbf{z}_i$ are called primal variables, $\mathbf{y}_i\in \mathbb{R}^n$ is the Lagrangian variable and  $\rho>0$ is called the penalty parameter. 
The function in (\ref{eq.ADMM-aug}) can be minimized by the ADMM steps  \cite{boyd}
\begin{align}
\mathbf{x}_i^{t+1}&:=\argmin_{\mathbf{x}_i} \bigg(f_i(\mathbf{x}_i)+\frac{\rho}{2}\Vert \mathbf{x}_i-\mathbf{z}_i^t+\mathbf{u}_i^t\Vert_2^2\bigg) \label{eq.ADMM3}\\
\mathbf{z}_i^{t+1}&:=\argmin_{\mathbf{z}_i} g\left(\sum_{i=1}^N \mathbf{z}_i \right)+ \frac{\rho}{2}\sum_{i=1}^N\Vert \mathbf{z}_i-\mathbf{u}_i^t-\mathbf{x}_i^{t+1}\Vert_2^2\label{eq.ADMM4}\\
\mathbf{u}_i^{t+1}&:=\mathbf{u}_i^t+\mathbf{x}_i^{t+1}-\mathbf{z}_i^{t+1}\label{eq.ADMM5}
\end{align}
 where $\mathbf{u}_i$ is called a dual variable (scaled version of Lagrangian variable, $\mathbf{u}_i=\frac{\mathbf{y}_i}{\rho}$).\\
The problems (\ref{eq.ADMM3}) and (\ref{eq.ADMM5}) can be solved in parallel for $i=1,\hdots,N$. 
Let $\mathbf{q}_i=\mathbf{u}_i^t+\mathbf{x}_i^{t+1}$. Then (\ref{eq.ADMM4}) can be rewritten as
\begin{align}
\min\quad &g\left(\sum_{i=1}^N \mathbf{z}_i \right)+\frac{\rho}{2}\sum_{i=1}^N \Vert \mathbf{z}_i-\mathbf{q}_i\Vert_2^2\label{eq.ADMM6}\\
\textrm{s.t}\quad & \bar{\mathbf{z}}=\frac{1}{N}\sum_{i=1}^N \mathbf{z}_i\nonumber
\end{align}
with fixed variable $\bar{\mathbf{z}}\in\mathbb{R}^n.$ The problem (\ref{eq.ADMM6}) has the following solution
\begin{align}
\mathbf{z}_i=\mathbf{q}_i+\bar{\mathbf{z}}-\bar{\mathbf{q}}, \quad \text{where} \quad \bar{\mathbf{q}}=\frac{1}{N}\sum_{i=1}^N \mathbf{q}_i=\bar{\mathbf{u}}^t+\bar{\mathbf{x}}^{t+1},\label{eq.ADMM7}
\end{align} 
 $$\bar{\mathbf{u}}^t= \frac{1}{N}\sum_{i=1}^N \mathbf{u}_i^t, \hspace{2mm}\text{and}\hspace{2mm} \bar{\mathbf{x}}^{t+1}= \frac{1}{N}\sum_{i=1}^N \mathbf{x}_i^{t+1}.$$
From (\ref{eq.ADMM6}) and (\ref{eq.ADMM7}), the $\bar{\mathbf{z}}$-update step can be simplified to the following unconstrained problem: 
\begin{align}
\min\quad g(N\bar{\mathbf{z}})+\frac{\rho}{2}\sum_{i=1}^N \Vert \bar{\mathbf{z}}-\bar{\mathbf{q}}\vert.
\end{align}
Substituting (\ref{eq.ADMM7}) for $\mathbf{z}_i^{t+1}$ into  (\ref{eq.ADMM5}), the $\bar{\mathbf{u}}$-update step is
\begin{align}
\mathbf{u}_i^{t+1}=\bar{\mathbf{u}}^t+\bar{\mathbf{x}}^{t+1}-\bar{\mathbf{z}}^{t+1}\label{16}
\end{align}
The dual variables $\{\mathbf{u}_i^t\}_{i=1}^N$ are equal and can be replaced by a single  variable $\mathbf{u}^t$.  The ADMM steps are simplified as follows 
\begin{align}
\mathbf{x}_i^{t+1}&:=\argmin_{\mathbf{x}_i} \bigg(f_i(\mathbf{x}_i)+\frac{\rho}{2}\Vert \mathbf{x}_i-\mathbf{x}_i^t+\bar{\mathbf{x}}^t-\bar{\mathbf{z}}^t+u^t\Vert_2^2\bigg)\label{eq16}\\
\bar{\mathbf{z}}^{t+1}&:=\argmin_{\bar{\mathbf{z}}} \bigg(g(N\bar{\mathbf{z}})+ \frac{N\rho}{2}\Vert \bar{\mathbf{z}}-u^t-\bar{\mathbf{x}}^{t+1}\Vert_2^2\bigg)\label{eq17}\\
\mathbf{u}^{t+1}&:=\mathbf{u}^t+\bar{\mathbf{x}}^{t+1}-\bar{\mathbf{z}}^{t+1}\label{eq18}
\end{align}
The original sharing problem (\ref{eq.ADMM1}) is decomposed into a three-step iterative optimization problem. The problem (\ref{eq16}) can be solved in parallel for $ i=1,\hdots, N$. The step (\ref{eq17}) solves the shared objective,  and (\ref{eq18}) is the dual variable update step.
\vspace{-1mm}
\section{ ADMM-based detection for UL CD-NOMA systems }  \label{sec.2}
This section focuses on reformulating the relation between the input and received signals of three multi-antenna CD-NOMA system models. These models are  1. SIMO-CD-NOMA, 2. Spatial multiplexing CD-NOMA (SMX-CD-NOMA), and 3. Spatial modulated CD-NOMA (SM-CD-NOMA). The reformulated models are readily applied for sharing ADMM-based detection process. These models, along with the derivation of the ADMM steps, are described in the following. 
\subsection{SIMO CD-NOMA}\label{SIMO_sub1}
Consider a UL scenario with the BS equipped with $N_\mathrm r$ antennas and  each UE having a single antenna. 
The idea is to exploit the spatial diversity gain offered by multiple receive antennas at BS \cite{marzetta}.  The error rate performance of the CD-NOMA system with $N_{\mathrm r}$ antennas at the receiver is significantly improved \cite{Lim} due to higher diversity gain.  The $K\times 1$ observation vector at $n_{\mathrm r}$th BS antenna can be expressed as
\begin{align*}
\mathbf{r}^{(n_{\mathrm r})}=\sum_{j=1}^J \text{diag}\left(\mathbf{h}_j^{(n_{\mathrm r})}\right)\mathbf{x}_j+\mathbf{w}^{(n_{\mathrm r})}, \hspace{2mm} n_{\rm r}=1,\hdots,N_{\rm r}
\end{align*}
where 
\begin{itemize}
        \item $\mathbf{w}^{(n_{\mathrm r})}\sim \mathcal{CN}(0,\sigma^2\mathbf{I}_K)$, $K\times 1$ vector assumed to be independent and identically distributed (i.i.d.),   AWGN at the $n_{\mathrm r}$th BS antenna.  
	\item $\mathbf{h}_j^{(n_{\mathrm r})}\sim \mathcal{CN}(0,\mathbf{I}_K)$, $K\times 1$ vector assumed to be i.i.d. complex Gaussian random vector,   denotes     Rayleigh  fading channel coefficients between the  $j$th UE and $n_{\mathrm r}$th BS antenna. 
	  $\mathbf{h}_j^{(n_{\mathrm r})}=\left[h[1]\ldots h[k] \ldots h[K]\right]^T$ and $\text{diag}\left(\mathbf{h}_j^{(n_{\mathrm r})}\right)$ denotes diagonal matrix with $h[k]$ being the $k$th diagonal element.
	  \item $\mathbf{x}_j=\left[x_j[1]\ldots x_j[k] \ldots x_j[K]\right]^T$   is a codeword from the  $j$th UE's codebook, $\mathcal{X}_j^{K\times M}$.
\end{itemize} 
The overall $KN_{\mathrm r}\times 1$ observation vector at the BS for the SIMO CD-NOMA system is given by
\begin{align}\label{eq1}
\mathbf{r}=\mathbf{H}\mathbf{x}_{\rm{mu}}+\mathbf{w},
\end{align} 
where 
\begin{itemize}
\item $\mathbf{r}=[\mathbf{r}^{(1)^T}\hdots\mathbf{r}^{(n_{\mathrm r})^T},\hdots, \mathbf{r}^{(N_{\mathrm r})^T}]^T$.
\item $\mathbf{w}\sim \mathcal{CN}(0,\sigma^2\mathbf{I}_{KN_{\mathrm r}}), N_{\mathrm r}K\times 1$ vector denotes the AWGN vector at the BS.
\item $\mathbf{x}_{\rm{mu}}$ is the $J N_{\mathrm e}\times 1$ multi-user concatenated transmitted signal.\\  $\mathbf{x}_{\rm{mu}}=[x_1[1]\hdots x_1[N_{\mathrm e}] \hdots x_j[1]\hdots x_j[N_{\mathrm e}]\hdots x_J[1] \hdots x_J[N_{\mathrm e}]]^T$, 
where $N_{\mathrm e}$ is the number of nonzero elements in codeword. We have $N_{\mathrm e}=K$ and $N_{\mathrm e}=d_{\mathrm v}$ for DCMA and SCMA, respectively. The transmitted signal $\mathbf{x}_{\rm{mu}}$ can be rewritten to facilitate the sharing-based detection problem in Section \ref{ADMM_sub} i.e., $\mathbf{x}_{\rm{mu}}=\sum_{j=1}^J \mathbf{x}_{0j}$.  The variable $\mathbf{x}_{0j}=[0 \hdots 0\hspace{2mm} x_{j}[1]\hspace{2mm} x_{j}[2]\hdots x_{j}[N_\mathrm e]\hspace{2mm} 0\hdots 0]^T$ represents the $j$th UE codeword.
		
\item  The channel matrix for DCMA system of size $KN_{\mathrm r} \times JK $  is given by
		\begin{align*}\mathbf{H}=
			\begin{bmatrix}
				\text{diag}(\mathbf{h}_1^{(1)})&\hdots&\text{diag}(\mathbf{h}_j^{(1)})&\hdots&\text{diag}(\mathbf{h}_J^{(1)})\\
				\text{diag}(\mathbf{h}_1^{(2)})&\hdots&\text{diag}(\mathbf{h}_j^{(2)})&\hdots&\text{diag}(\mathbf{h}_J^{(2)})	
				\\	\vdots&\hdots&\vdots&\hdots&\vdots\\
			\text{diag}(\mathbf{h}_1^{(N_{\mathrm r})})&\hdots&\text{diag}(\mathbf{h}_j^{(N_{\mathrm r})})&\hdots&\text{diag}(\mathbf{h}_J^{(N_{\mathrm r})})									
			\end{bmatrix}.    
		\end{align*}
\item The channel matrix for SCMA system of size $KN_{\mathrm r} \times Jd_{\mathrm v} $  is given by
\begin{align*}\mathbf{H}=
			\begin{bmatrix}
				\overline{\text{diag}(\mathbf{h}_1^{(1)})}&\hdots&\overline{\text{diag}(\mathbf{h}_j^{(1)})}&\hdots&\overline{\text{diag}(\mathbf{h}_J^{(1)})}\\
				\overline{\text{diag}(\mathbf{h}_1^{(2)})}&\hdots&\overline{\text{diag}(\mathbf{h}_j^{(2)})}&\hdots&\overline{\text{diag}(\mathbf{h}_J^{(2)})}\\	\vdots&\hdots&\vdots&\hdots&\vdots\\
			\overline{\text{diag}(\mathbf{h}_1^{(N_{\mathrm r})})}&\hdots&\overline{\text{diag}(\mathbf{h}_j^{(N_{\mathrm r})})}&\hdots&\overline{\text{diag}(\mathbf{h}_J^{(N_{\mathrm r})})}									
			\end{bmatrix}.  
		\end{align*}
		where $\overline{\text{diag}\left(\mathbf{h}_j^{(N_{\mathrm r})}\right)}$ is $K\times d_{\mathrm v}$ matrix after removing the columns in $\text{diag}(\mathbf{h}_j^{(N_{\mathrm r})})$ corresponding to zero elements of $j$th UE codeword $\mathbf{x}_j$ (inactive REs of $j$th UE).
		
\end{itemize}
\begin{ex}
    Consider the factor graph matrix for the 150~\% overloading SCMA system with  $K=4, J=6, d_{\mathrm f}=3, d_{\mathrm v}=2$:
\vspace{-0.2in}
\begin{align}\label{FG}
\mathbf{F}=
			\begin{bmatrix}
   1 & 0 & 1 & 0 & 1 & 0 \\
0 & 1 & 1 & 0 & 0 & 1\\
 1 & 0 & 0 & 1 & 0 & 1\\
 0 & 1 & 0 & 1 & 1 & 0
                \end{bmatrix}. 
\end{align}
For the $1$st UE and the $n_{\mathrm r}$th receiving antenna, the channel matrix is given below:
\[\overline{\rm{diag}(\mathbf{h}_1^{(n_{\mathrm r})})}=\left[
\begin{tabular}{cccccc}
$h_1^{(n_{\mathrm r})}[1]$&0 \\
0&0\\
0&$h_1^{(n_{\mathrm r})}[3]$\\
0&0

\end{tabular}
\right].
\]

The transmitted signal from all UEs is given by 
\begin{align} \label{xmu}  
{{\bf{x}}_{{\rm{mu}}}} = {\left[ {\underbrace {{x_1}[1]\;\;{x_1}[3]}_{{\rm{UE}}\;1}\;\;\underbrace {{x_2}[2]\;\;{x_2}[4]}_{{\rm{UE}}\;2}\;\;\underbrace {{x_3}[1]\;\;{x_3}[2]\;}_{{\rm{UE}}\;3}\;\underbrace {{x_4}[3]\;\;{x_4}[4]}_{{\rm{UE}}\;4}\;\;\underbrace {{x_5}[1]\;\;{x_5}[4]\;}_{{\rm{UE}}\;5}\;\underbrace {{x_6}[2]\;\;{x_6}[3]}_{{\rm{UE}}\;6}} \right]^T}.
\end{align}
The 3rd UE codeword $\mathbf{x}_{03}$ in ${{\bf{x}}_{{\rm{mu}}}}$ is given by
\begin{align} \label{xmu2}  
{\mathbf{x}_{03}} = {\left[ {\underbrace {{0}\;\;{0}}_{{\rm{UE}}\;1}\;\;\underbrace {{0}\;\;{0}}_{{\rm{UE}}\;2}\;\;\underbrace {{x_3}[1]\;\;{x_3}[2]\;}_{{\rm{UE}}\;3}\;\underbrace {{0}\;\;{0}}_{{\rm{UE}}\;4}\;\;\underbrace {{0}\;\;{0}\;}_{{\rm{UE}}\;5}\;\underbrace {{0}\;\;{0}}_{{\rm{UE}}\;6}} \right]^T}.
\end{align}
The variables $\{\mathbf{x}_{0j}\}_{j=1}^J$ are similarly represented as $\mathbf{x}_{03}$. Note that, the transmitted signal $\mathbf{x}_{\rm{mu}}$ can be written as, $\mathbf{x}_{\rm{mu}}=\sum_{j=1}^6\mathbf{x}_{0j}$.
\label{ex1}
\hfill \qedsymbol
\end{ex} 
\vspace{-3mm}
\subsection{SMX-CD-NOMA}\label{smx_sub}
\begin{figure}[!htbp]
\centering
\includegraphics[width=17cm, height=8cm]{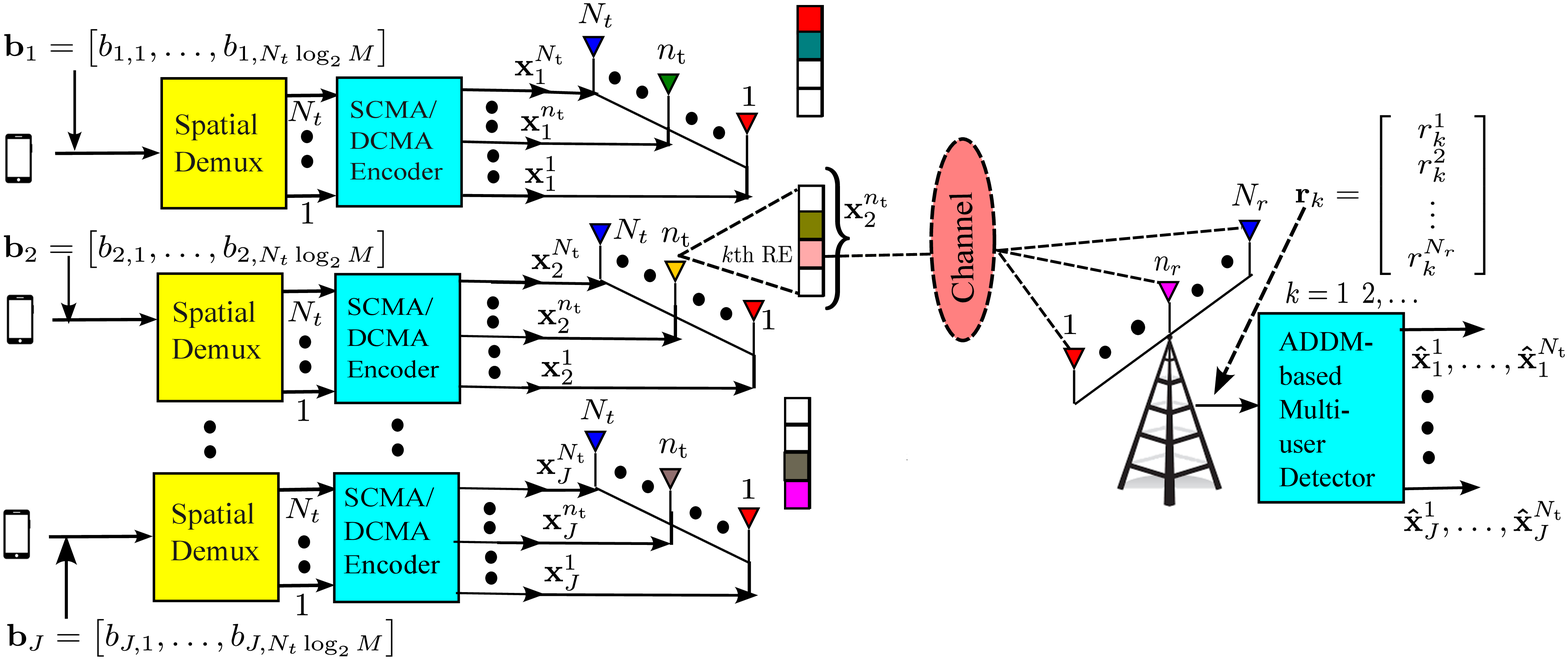}    
\caption{SMX-CD-NOMA system model in the UL. }\label{SMX_new}
\end{figure}

\begin{figure}[h]
\centering
\includegraphics[scale=0.4]{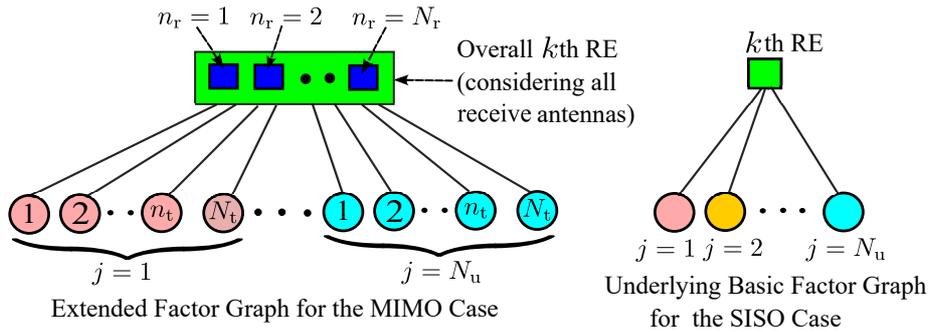}      
\caption{Processing at $k$th RE of SMX-CD-NOMA system.}\label{SMX}
\end{figure}

The spectral efficiency of the UL CD-NOMA system is further enhanced by placing multiple antennas at the transmitter \cite{Elkawafi}. Multiple antennas at both transmitter and  receiver of the CD-NOMA system are considered in this section. Multiple antennas at the transmitter exploit multiplexing gain offered by spatially multiplexing several data streams onto the MIMO channel.
Consider a scenario where each UE is equipped with $N_{\mathrm t}$ transmitting antennas, and the BS is equipped with $N_{\mathrm r}$ receiving antennas. The total input data at each UE, $N_{\mathrm t} \text{log}_2(M)$ bits, are divided into $N_{\mathrm t}$ parallel data streams. Each data stream is fed to the CD-NOMA encoder, as shown in Fig. \ref{SMX_new}. Note that in the SMX-CD-NOMA system, all the UE antennas transmit data simultaneously. The main challenge of SMX-CD-NOMA in practical implementation is the computational complexity at both transmitter and receiver. Further, detecting the data transmitted from multiple antennas of multiple UEs is a complex operation. The proposed method performs resource-wise processing at the BS via the ADMM algorithm, as shown in Fig. \ref{SMX}.  

The received signal at the BS over the $k$th RE is given by
\begin{align}\label{eq2}
\mathbf{r}_k=\mathbf{H}_k\mathbf{x}_{\rm{smx},k}+\mathbf{w}_k,
\end{align} 
where
\begin{itemize}
\item $\mathbf{r}_k=[r_k^1\hdots r_k^{n_{\mathrm r}}\hdots r_k^{N_{\mathrm r}}]^T$, $N_{\mathrm r}\times 1$ observation vector.
\item  $\mathbf{w}_k\sim \mathcal{CN}(0,\sigma^2\mathbf{I}_{N_{\mathrm r}}), N_{\mathrm r}\times 1$ vector denotes the AWGN at the BS over $k$th RE.
\item $\mathbf{x}_{{\rm{smx}},k}$ is the  $N_{\mathrm u}N_{\mathrm t}\times 1$ transmitted vector on $k$th RE. $\mathbf{x}_{{\rm{smx}},k}=[\mathbf{x}_{1,k}^T\hdots\mathbf{x}_{j,k}^T\hdots\mathbf{x}_{N_{\mathrm u},k}^T]^T$,
 and each  $\mathbf{x}_{j,k}=[x_{j,k,1}\hdots x_{j,k,n_{\mathrm t}} \hdots x_{j,k,N_{\mathrm t}}]^T$ is $N_{\mathrm t}\times 1$ vector corresponding to $j$th UE. Each $x_{j,k,n_{\mathrm t}}\in \mathbf{x}_{ j}^{n_{\mathrm t}}$ is the symbol transmitted from $n_{\mathrm t}$th antenna of  $j$th UE over $k$th RE, where  $\mathbf{x}_{ j}^{n_{\mathrm t}}$ is the codeword of $j$th UE transmitted from $n_{\mathrm t}$th antenna.
The set $\zeta_k$ represents the set of UEs overlapping on $k$th RE given by 
 $$\zeta_k=\lbrace j: \mathbf{x}_j[k]\neq 0;1\leq j\leq J\rbrace, \hspace{2mm} \text{and}\hspace{2mm} \vert\zeta_k\vert= N_{\mathrm u},$$
where  $ N_{\mathrm u}=d_{\mathrm f}$ and $ N_{\mathrm u}=J$ for SCMA and  DCMA, respectively.

The transmitted signal $\mathbf{x}_{{\rm{smx}},k}$ can be rewritten to formulate the sharing-based detection problem in Section \ref{ADMM_sub} i.e., $\mathbf{x}_{{\rm{smx}},k}= \sum_{j=1}^{N_{\mathrm u} }\mathbf{x}_{0j,k}$. The variable $\mathbf{x}_{0j,k}=[0\hdots 0\hspace{2mm} x_{j,k,1} \hspace{2mm}x_{j,k,2}\hdots x_{j,k,N_{\rm t}}\hspace{2mm}0\hdots 0]^T$ is the  $N_{\mathrm u}N_{\mathrm t}\times 1$ vector, and represents the $j$th overlapped UE on $k$th RE. The nonzero elements in $\mathbf{x}_{0j,k}$ are symbols transmitted from $N_{\rm t}$ antennas.

\item $\mathbf{H}_k$ is the  $N_{\mathrm r}\times N_{\mathrm u}N_{\mathrm t}$ matrix given by
$$\mathbf{H}_k=[\mathbf{H}_{1,k}\hdots\mathbf{H}_{j,k}\hdots\mathbf{H}_{N_{\mathrm u},k}]$$
where $\mathbf{H}_{j,k}$ represents the $j$th UE MIMO channel matrix of size $N_{\mathrm r}\times N_{\mathrm t}$.



\end{itemize}

\begin{ex}

Consider $N_{\mathrm t}=2$, and according to the factor graph matrix given in (\ref{FG}). The transmitted  signal over the first RE in the SCMA system is 
\begin{equation}\label{eq_smx}
\mathbf{x}_{\rm{smx},1} = 
\begin{bmatrix}
\mathbf{x}_{1,1}^T&\mathbf{x}_{3,1}^T&\mathbf{x}_{5,1}^T
\end{bmatrix}^T.
\end{equation}
For DCMA, due to the dense structure of codebooks, all $J$ UEs  overlap on each RE \cite{Liu}. The transmitted signal over the first RE is given by 
\begin{equation}\label{eq_smx_dcma}
\mathbf{x}_{\rm{smx},1} = 
\begin{bmatrix}
\mathbf{x}_{1,1}^T & \mathbf{x}_{2,1}^T&\hdots & \mathbf{x}_{5,1}^T&\mathbf{x}_{6,1}^T
\end{bmatrix}^T,
\end{equation}
where each $\mathbf{x}_{j,1}$ is a $2\times 1$ ($N_{\mathrm t}\times 1$) vector  given by
\begin{equation*}
\mathbf{x}_{j,1} = 
\begin{bmatrix}
x_{j,1,1}&x_{j,1,2}
\end{bmatrix}^T.
\end{equation*}

The transmitted signal over each remaining RE ($\mathbf{x}_{{\rm{smx}},k}, k=2,\hdots,K$) has a similar representation as in (\ref{eq_smx}) and (\ref{eq_smx_dcma}) for SCMA and DCMA, respectively. 

For SCMA,   $\mathbf{x}_{03,1}$ represents the 3rd UE's symbols as given by
\begin{align} \label{xmu2}  
{\mathbf{x}_{03,1}} = {\left[ {\underbrace {{0}\;\;{0}}_{{\rm{UE}}\;1}\;\;\underbrace {{x_{3,1,1}}\;\;{x_{3,1,2}}\;}_{{\rm{UE}}\;3}\;\;\underbrace {{0}\;\;{0}\;}_{{\rm{UE}}\;5}} \right]^T}.
\end{align}
For DCMA,   $\mathbf{x}_{03,1}$ can be written as
\begin{align} \label{xmu2}  
{\mathbf{x}_{03,1}} = {\left[ {\underbrace {{0}\;\;{0}}_{{\rm{UE}}\;1}\;\;\underbrace {{0}\;\;{0}}_{{\rm{UE}}\;2}\;\;\underbrace {{x_{3,1,1}}\;\;{x_{3,1,2}}\;}_{{\rm{UE}}\;3}\;\underbrace {{0}\;\;{0}}_{{\rm{UE}}\;4}\;\;\underbrace {{0}\;\;{0}\;}_{{\rm{UE}}\;5}\;\underbrace {{0}\;\;{0}}_{{\rm{UE}}\;6}} \right]^T}.
\end{align}
The variables $\{\mathbf{x}_{0j,k}\}_{k=1}^K$ are similarly represented as $\mathbf{x}_{03,1}$. The transmitted signal $\mathbf{x}_{{\rm{smx}},k}$ can be written as, $\mathbf{x}_{{\rm{smx}},k}= \sum_{j=1}^3 \mathbf{x}_{0j,k}$  and $\mathbf{x}_{{\rm{smx}},k}= \sum_{j=1}^6 \mathbf{x}_{0j,k}$ for SCMA and DCMA, respectively.
\label{ex2}
\hfill 
\qedsymbol
\end{ex}
\subsection{SM-CD-NOMA} 
\begin{figure}[!htbp]
\centering
\includegraphics[scale=0.45]{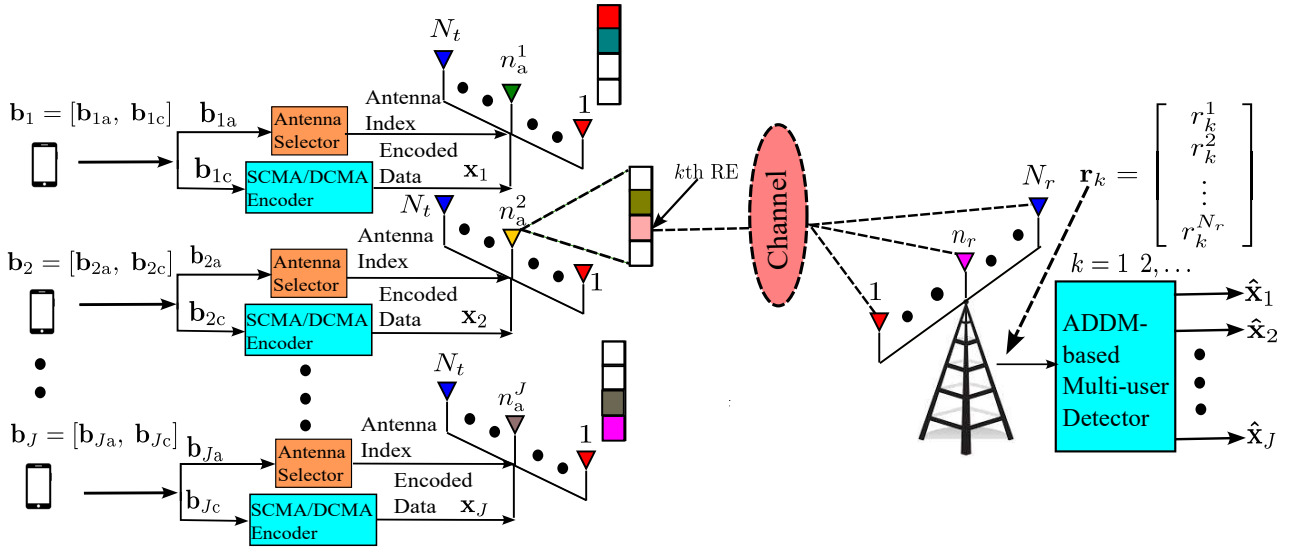}
\caption{ SM-CD-NOMA system model in the UL.}\label{SMX_}
\end{figure}

Due to the requirement of  multiple radio frequency (RF) chains,  the SMX-CD-NOMA system is not affordable for various applications. Further,  inter-channel interference is the main limitation of this system. Spatial modulation (SM) MIMO systems are promising technology to overcome the limitations of SMX MIMO systems. SM for single-UE communications is well studied in the literature \cite{Mesleh}. Due to the demands of future wireless networks, it is important to study and analyze SM in multiuser scenarios. In \cite{Yusha}, the authors studied and analyzed SM sparse CDMA. SM-aided CD-NOMA (SM-CD-NOMA) is guaranteed to improve the spectral efficiency  of the CD-NOMA with feasible complexity at both transmitter and receiver \cite{Pan, Yaoyue}. Fig. \ref{SMX_} shows the system model for the SM-CD-NOMA system. Each UE is equipped with $N_{\mathrm t}$ transmitting antennas, in which only one antenna is active at any time, as shown in Fig.~\ref{SMX_}. The active antenna index of the $j$th UE is denoted by  $n_{\rm a}^j$. All other antennas remain silent in that particular time slot. Thus, the active antenna index is a spatial modulation symbol to transmit extra information bits. The information bit stream $\mathbf{b}_j$ of the $j$th UE,  is split into two parts [$\mathbf{b}_{j\mathrm a}$,$\mathbf{b}_{j\mathrm c}$] having $\text{log}_2(N_{\mathrm t})$ and $\text{log}_2(M)$ bits, respectively.   Each UE  transmits an overall $\text{log}_2(N_{\mathrm t})+\text{log}_2(M)$  bits from the active antenna.

\begin{figure}[!htbp]
\centering
\includegraphics[scale=0.4]{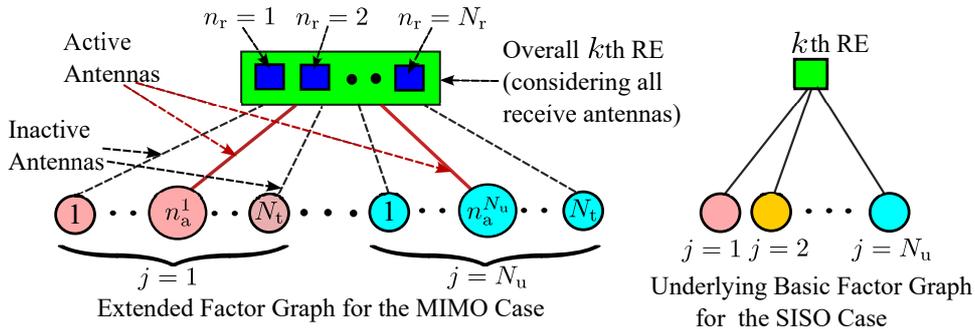}      
\caption{Processing at $k$th RE of SM-CD-NOMA system.}\label{fig:SM}
\end{figure}

The observation vector over the $k$th RE at the BS is an $N_{\mathrm r}\times 1$ vector  given by
\begin{align}\label{eq3}
\mathbf{r}_k=\mathbf{H}_k\mathbf{x}_{\text{sm},k}+\mathbf{w}_k,
\end{align}
where
\begin{itemize}
\item  $\mathbf{r}_k,\mathbf{w}_k$, and $\mathbf{H}_k$ have the similar forms as in  SMX-CD-NOMA for both DCMA and SCMA systems. 

\item  $\mathbf{x}_{\text{sm},k} $ is the  $N_{\mathrm u}N_{\mathrm t}\times 1$ vector 
$\mathbf{x}_{\text{sm},k} =[\mathbf{x}_{1,k}^T\hdots\mathbf{x}_{j,k}^T\hdots\mathbf{x}_{N_{\mathrm u},k}^T]^T$,
 and each  $\mathbf{x}_{j,k}=[0\hdots x_{j,k,n_{\rm a}^j} \hdots 0]^T$ is $N_{\mathrm t}\times 1$ vector corresponding to $j$th UE. The nonzero element $x_{j,k,n_{\rm a}^j}$ is the symbol transmitted from $n_{\rm a}^j$th active antenna.
 The transmitted signal $\mathbf{x}_{\text{sm},k}$ can be rewritten to formulate the sharing-based detection problem in Section \ref{ADMM_sub} i.e., $\mathbf{x}_{\text{sm},k}= \sum_{j=1}^{N_{\mathrm u} }\mathbf{x}_{0j,k}^{n_{\rm a}^j}$. The variable $\mathbf{x}_{0j,k}^{n_{\rm a}^j}=[0\hdots0\hdots\hspace{2mm} x_{j,k,n_{\rm a}^j}\hspace{2mm}0\hdots\hspace{2mm}0\hdots0]^T$ represents the $j$th overlapped UE on $k$th RE. The nonzero element in $\mathbf{x}_{0j,k}^{n_{\rm a}^j}$  is the symbol transmitted from $n_{\rm a}^j$th active antenna.

\end{itemize}
 Fig. \ref{fig:SM} depicts the $k$th RE processing at the BS through the extended factor graph. The thick and dotted lines correspond to the  active and inactive antennas, respectively. 

\begin{ex}
Consider $N_{\mathrm t}=2$, and the factor graph matrix given in (\ref{FG}). Suppose the active antenna indices of the $6$-users are \{2,1,2,2,1,1\}.
The structure of transmit codewords from $J$ UEs for SM-CD-NOMA is as follows \cite{Yaoyue}:
\begin{align}
\mathbf{x}_{\rm{tx}} =
			\begin{bmatrix}
					\mathbf{0}_	{K\times 1}&\mathbf{x}_{2_{K\times 1}}^1&	\mathbf{0}_	{K\times 1}&\mathbf{0}_	{K\times 1}&\mathbf{x}_{5_{K\times 1}}^1&\mathbf{x}_{6_{K\times 1}}^1\\
                   \mathbf{x}_{1_{K\times 1}}^2& \mathbf{0}_	{K\times 1}&\mathbf{x}_{3_{K\times 1}}^2&\mathbf{x}_{4_{K\times 1}}^2&\mathbf{0}_	{K\times 1}&\mathbf{0}_	{K\times 1}\\
			\end{bmatrix}_{N_{\mathrm t}K\times J =8\times 6} .
   \label{eq::sm_tx}
		\end{align}
  Each $\mathbf{x}_{j_{_{K\times 1}}}^{n_{\rm a}^j}$ in $\mathbf{x}_{\rm{tx}}$ represents the transmitted codeword from $j$th UE, and each column of $\mathbf{x}_{\text{tx}}$ carries information about active transmit antenna ($n_{\rm a}^j$) as well as the codeword. In (\ref{eq::sm_tx}), $\mathbf{0}_	{K\times 1}$ indicates zero power transmitted by a deactivated antenna.
To facilitate the resource-wise processing at the receiver via ADMM, the transmitted signal over the first RE in the SCMA system is modeled as
\begin{equation}\label{re1_SCMA}
\mathbf{x}_{\rm{sm},1} = 
\begin{bmatrix}
\mathbf{x}_{1,1}^T&\mathbf{x}_{3,1}^T&\mathbf{x}_{5,1}^T\hspace{3mm}
\end{bmatrix}^T.
\end{equation}
The transmitted signal over the first RE in the DCMA system is modeled as
\begin{equation}\label{re1}
\mathbf{x}_{\rm{sm},1} = 
\begin{bmatrix}
\mathbf{x}_{1,1}^T&\mathbf{x}_{2,1}^T&\hdots&\mathbf{x}_{5,1}^T&\mathbf{x}_{6,1}^T\hspace{3mm}
\end{bmatrix}^T,
\end{equation}
where   $\mathbf{x}_{1,1}=[0\hspace{2mm} x_{1,1,2}]^T$,  $\mathbf{x}_{2,1}=[x_{2,1,1}\hspace{2mm} 0]^T$, $\mathbf{x}_{3,1}=[0\hspace{2mm} x_{3,1,2}]^T$,  $\mathbf{x}_{4,1}=[0\hspace{2mm} x_{4,1,2}]^T$,  $\mathbf{x}_{5,1}=[x_{5,1,1}\hspace{2mm} 0]^T$, and  $\mathbf{x}_{6,1}=[x_{6,1,1} \hspace{2mm}0]^T$. The transmitted signals over the remaining REs ($\mathbf{x}_{{\rm{sm}},k}, k=2,\hdots,K$ ) have similar representations as (\ref{re1_SCMA}) and (\ref{re1}) for SCMA and DCMA, respectively.
For SCMA,   $\mathbf{x}_{03,1}^2$ represents the symbol transmitted from 2nd active antenna of 3rd UE on 1st RE. $\mathbf{x}_{03,1}^2$  can be written as 
\vspace{-0.3mm}
\begin{align} \label{xmu2}  
{\mathbf{x}_{03,1}^2} = {\left[ {\underbrace {{0}\;\;{0}}_{{\rm{UE}}\;1}\;\;\underbrace {{0}\;\;{x_{3,1,2}}\;}_{{\rm{UE}}\;3}\;\;\underbrace {{0}\;\;{0}\;}_{{\rm{UE}}\;5}} \right]^T}.
\end{align}
For DCMA,   $\mathbf{x}_{03,1}^2$ can be written as
\begin{align} \label{xmu2}  
{\mathbf{x}_{03,1}^2} = {\left[ {\underbrace {{0}\;\;{0}}_{{\rm{UE}}\;1}\;\;\underbrace {{0}\;\;{0}}_{{\rm{UE}}\;2}\;\;\underbrace {{0}\;\;{x_{3,1,2}}\;}_{{\rm{UE}}\;3}\;\underbrace {{0}\;\;{0}}_{{\rm{UE}}\;4}\;\;\underbrace {{0}\;\;{0}\;}_{{\rm{UE}}\;5}\;\underbrace {{0}\;\;{0}}_{{\rm{UE}}\;6}} \right]^T}.
\end{align}
The variables $\{\mathbf{x}_{0j}^{n_{\rm a}^j}\}_{k=1}^K$ are similarly represented as $\mathbf{x}_{03,1}^2$. The transmitted signal $\mathbf{x}_{{\rm{sm}},k}$ can be written as, $\mathbf{x}_{{\rm{sm}},k}= \sum_{j=1}^3 \mathbf{x}_{0j,k}^{n_{\rm a}^j}$  and $\mathbf{x}_{{\rm{sm}},k}= \sum_{j=1}^6 \mathbf{x}_{0j,k}^{n_{\rm a}^j}$ for SCMA and DCMA, respectively
\label{ex3}
\hfill \qedsymbol
\end{ex}
Observe from equations (\ref{eq1}), (\ref{eq2}), and (\ref{eq3}) that the  system models of three MIMO-CD-NOMA systems are similar to those of the conventional MIMO system model. 
This model is given by
\begin{align}\label{eq.ML-mimo}
\mathbf{r}=\mathbf{H}\mathbf{x}+\mathbf{w},~ \mathbf{x}\in \mathcal{X}^{N_{\mathrm t}}, 
\end{align}
where $\mathcal{X}$ is the signal constellation, $\mathbf{r}$ is the $N_{\mathrm r}\times 1$ observation vector, $\mathbf{H}$ is $N_{\mathrm r}\times N_{\mathrm t}$ channel matrix ($N_{\mathrm r}>N_{\mathrm t}$), $\mathbf{x}$ is $N_{\mathrm t}\times 1$ transmitted vector and $\mathbf{w}$ is i.i.d.  AWGN vector with each component being distributed as $\mathcal{CN}(0,\sigma^2)$. 
 The ML detection problem can be formulated as
\begin{align*}
\text{min}_{\mathbf{x}\in \mathcal{X}^{N_{\rm t}}} \Vert\mathbf{r}-\mathbf{H}\mathbf{x}\Vert^2. 
\end{align*} 
The ML detection problem for SIMO CD-NOMA system in (\ref{eq1}) is given by
\begin{align}\label{eq5}
\min_{\mathbf{x}_{\rm{mu}}\in\mathcal{X}_{\text{mu}}^{JN_{\rm e}} }\quad \Vert \mathbf{r}-\mathbf{H}\mathbf{x}_{\rm{mu}}\Vert^2, 
\end{align}
where $\mathcal{X}_{\text{mu}}^{JN_{\rm e}} $ denotes the multi-user signal constellation and it consists of $J$-UEs concatenated codewords. However, solving ML detection problems, in general, is NP-hard. The exhaustive search method can be used to solve the ML detection problem. However, the exhaustive search is exponentially complex as per $\mathcal{O}(M^J)$  \cite{Yang} and  thus it  is not feasible. The ML detection problem (\ref{eq5}) can be solved with a polynomial complexity using a distributed optimization framework. The above problem needs to be converted into a sharing problem that can be solved using distributed optimization methods. In the next section, we apply an efficient method based on the ADMM algorithm to solve the sharing-based detection problem in a distributed manner.

\subsection{Large-scale UL multi-antenna CD-NOMA detection via ADMM}\label{ADMM_sub}
This section discusses the design of ADMM-based MUD for two CD-NOMA techniques, i.e., SCMA and DCMA. 
 The ADMM-based detector is applied to three different MIMO systems, namely, SIMO-CD-NOMA, spatial multiplexed CD-NOMA (SMX-CD-NOMA), and spatially modulated CD-NOMA (SM-CD-NOMA). The implications of the proposed method are discussed in subsequent sections.   
\subsubsection{SIMO CD-NOMA }
The ML detection problem in (\ref{eq5}) can be converted into a sharing problem. Further, this problem can be solved in a distributed manner via the ADMM approach, as discussed in  Section~\ref{sec.II(C)}. ADMM allows parallel processing in MUD problems, guaranteeing a minimal computational time at the receiver. 
 Recall  from Section \ref{SIMO_sub1}, the  transmitted signal $\mathbf{x}_{\rm{mu}}$ is given by $\mathbf{x}_{\rm{mu}}=\sum_{j=1}^J \mathbf{x}_{0j}$.  
 The  problem (\ref{eq5}) can be rewritten as 
 \vspace{-2mm}
\begin{align}
\min_{\mathbf{x}_{0j}}\quad& \Vert \mathbf{r}-\mathbf{H}\left(\sum_{j=1}^J \mathbf{x}_{0j}\right)\Vert^2\label{eq19}\\
\textrm{s.t}\quad & \mathbf{x}_{0j}\in \mathcal{X}_{\text{mu}}^{JN_{\mathrm e}} \quad  j=1,\hdots,J.\nonumber
\end{align}
Here, for the channel matrix $\mathbf{H}^{N_{\mathrm r}K\times JN_{\mathrm e}}$, we consider $N_{\mathrm r}K>JN_{\mathrm e}$. 
Each entry's real and imaginary parts in $\mathbf{x}_{0j}$ are restricted by set constraints. The set constraint  must be converted into an interval constraint to convert  (\ref{eq19}) into a sharing problem.
Box constraint relaxation (BCR) is used to relax the set constraints of each entry in the codeword\cite{Zhang_Quan}. Hence, each entry's real and imaginary parts in the $j$th UE codeword belong  to $[-\alpha_j,\alpha_j]$ and $[-\beta_j,\beta_j]$.  After relaxation, each element in the $j$th UE codebook is defined as $\tilde{\mathcal{X}}_j=\{x_j=x_{jR}+ix_{jI}\vert x_{jR}\in [-\alpha_j,\alpha_j],x_{jI}\in[-\beta_j,\beta_j] \}$, $\alpha_j=\max\vert{\mathbb{R}(\mathcal{X}_{j})}\vert,\beta_j=\max\vert{\mathcal{I}(\mathcal{X}_{j})}\vert$. The highly complex MIMO-CD-NOMA ML detection problem in (\ref{eq19}) is now ready to be converted into a non-convex distributed optimization problem. However, the constraint relaxation degrades the detection performance due to the losses introduced by the interval constraints. The losses can be compensated by adding a set of quadratic penalty functions $\sum_{j=1}^J \frac{\gamma_j}{2}\Vert \mathbf{x}_{0j}\Vert_2^2$ to the objective function where $\gamma_j\geq 0$ is a penalty parameter. The penalty functions are selected so that each variable, $\mathbf{x}_{0j}$, in the penalty term  minimizes the individual penalty and the shared objective. The added penalty term makes the solution as sparse as possible. However, the sparse vectors with specific numbers ($d_{\mathrm v}$ for SCMA and $K$ for DCMA) of non-zero entries only will minimize the shared objective function.
The favorable solutions are the sparse vectors with $d_{\mathrm v}$ and $K$ non-zero elements for SCMA and DCMA, respectively.  The sharing ADMM problem can be written as  
\begin{align}
\min_{\mathbf{x}_{0j}}\quad&  \Vert \mathbf{r}-\mathbf{H}\left(\sum_{j=1}^J \mathbf{x}_{0j}\right)\Vert^2+\sum_{j=1}^J \frac{\gamma_j}{2}\Vert \mathbf{x}_{0j}\Vert_2^2 \label{eq.sh1-MU}\\
\textrm{s.t}\quad & \mathbf{x}_{0j}\in \tilde{\mathcal{X}}_{\text{mu}}^{JN_{\mathrm e}}, \quad  j=1,\hdots,J.\nonumber
\end{align}

The problem in (\ref{eq.sh1-MU}) is similar to the sharing ADMM problem in (\ref{eq.ADMM1}). Proceeding similarly as in Section \ref{sec.II(C)}, (\ref{eq.sh1-MU}) can be written as
\begin{align}
\min_{\mathbf{x}_{0j},\mathbf{z}_{0j}}\quad&  \Vert \mathbf{r}-\mathbf{H}\left(\sum_{j=1}^J \mathbf{x}_{0j}\right)\Vert^2+\sum_{j=1}^J \frac{\gamma_j}{2}\Vert \mathbf{z}_{0j}\Vert_2^2\label{eq.sh2-MU}\\
\textrm{s.t} \quad &\mathbf{x}_{0j}=\mathbf{z}_{0j},
\quad\mathbf{x}_{0j}\in \tilde{\mathcal{X}}_{\text{mu}}^{JN_{\mathrm e}}, \quad \forall j=1,\hdots,J\nonumber
\end{align}

For the formulation of the ADMM  steps,  the augmented Lagrangian function for the  problem (\ref{eq.sh2-MU}) is considered as shown below: 
\begin{align}
\mathcal{L}\left(\{\mathbf{x}_{0j},\mathbf{z}_{0j},\textbf{y}_j\}_{j=1}^J\right)=\Vert \textbf{r}-\mathbf{H}\left(\sum_{j=1}^J \mathbf{x}_{0j}\right)\Vert_2^2+\sum_{j=1}^J \frac{\gamma_j}{2}\Vert \mathbf{z}_{0j}\Vert_2^2\label{eq.aug}\\
+\sum_{j=1}^J {\text{Re}}\innerproduct{\mathbf{x}_{0j}-\mathbf{z}_{0j}}{\textbf{y}_j}+\frac{\rho}{2}\sum_{j=1}^J\Vert \mathbf{x}_{0j}-\mathbf{z}_{0j}\Vert_2^2\nonumber
\end{align}
where $\textbf{y}_j\in \mathbb{C}^{JN_{\mathrm e}}$ is Lagrangian multiplier of the $j$th UE. 
Letting $\mathbf{u}_j=\frac{\textbf{y}_j}{\rho}$, the scaled form of ADMM steps are as follows \cite{boyd} :
\begin{align}
\mathbf{z}_{0j}^{t+1}:=&\argmin_{ \mathbf{z}_{0j}\in \tilde{\mathcal{X}}_{\text{mu}}^{JN_{\rm e}}} \frac{\gamma_j}{2}\Vert \mathbf{z}_{0j}\Vert_2^2+\frac{\rho}{2}\Vert \mathbf{z}_{0j}-\mathbf{x}_{0j}^t+\mathbf{u}_{j}^t\Vert_2^2\label{eq.MU-ADMM1}\\
\mathbf{x}_{0j}^{t+1}:=&\argmin_{ \mathbf{x}_{0j}\in \tilde{\mathcal{X}}_{\text{mu}}^{JN_{\rm e}} }  \Vert \mathbf{r}-\mathbf{H}\left(\sum_{j=1}^J \mathbf{x}_{0j}\right)\Vert^2+\frac{\rho}{2}\sum_{j=1}^J\Vert \mathbf{z}_{0j}^{t+1}-\mathbf{x}_{0j}+\mathbf{u}_j^k\Vert_2^2\label{eq.MU-ADMM2}\\
\mathbf{u}_j^{t+1}:=&\mathbf{u}_j^k+(\mathbf{z}_{0j}^{t+1}-\mathbf{x}_{0j}^{t+1})\label{eq.MU-ADMM3}
\end{align}

\begin{algorithm}[!htbp]
  \small
 	\caption{SIMO CD-NOMA detection via sharing ADMM}
 	\begin{algorithmic}[1]
 	\State  \textbf{Input:} Noise variance ($N_0$), average codebook energy ($E_{\mathrm s}$) and initialize   $\lbrace\mathbf{z}_{0j}\rbrace_{j=1}^J, \bar{\mathbf{x}}_0,\mathbf{u},\bar{\mathbf{z}}_0$ with zero vectors.  \\	
 	 \textbf{Output:} $\bar{\mathbf{x}}_{0}^{(T)}$, where $T$ is the maximum number of iterations.
 	\State From (\ref{eq.MU-ADMM4}) and (\ref{eq.MU-ADMM5}),  obtain optimal solutions (\ref{eq.MU-ADMM8}) and (\ref{eq.MU-ADMM9}), respectively.
            \State \textbf{Preprocessing}\\
            $\rho=N_0/E_{\mathrm s}$\\
            $\mathbf{G}=(\mathbf{H}^H\mathbf{H}J+\rho\mathbf{I})^{-1}$\\
           $\mathbf{M}=\mathbf{H}^H\mathbf{r}$ 
  
			\For {$t=1,2,,\hdots T $}
 	\State \textbf{Step}:1 Update $\lbrace \mathbf{z}_{0j}^{t+1}\rbrace_{j=1}^J$  in parallel  via (\ref{eq.MU-ADMM8}) and compute $\bar{\mathbf{x}_1}$.
        \State \textbf{Step}:2
        
 	\State Update $\bar{\mathbf{x}}_0^{t+1}$ via (\ref{eq.MU-ADMM9}).
	\State Update $\mathbf{u}^{t+1}$ via (\ref{eq.MU-ADMM6})
		\EndFor
        \State Compute $\hat{\mathbf{x}}_{\text{mu}}=J\bar{\mathbf{x}}_0^{(T)}$. 
        \State After extracting each UE codeword from  $\hat{\mathbf{x}}_{\text{mu}}$, apply the MED rule according to (\ref{eq.MU-ADMM7}).

 	\end{algorithmic} 
  \label{alg1}
 \end{algorithm}
 \normalfont
 The variables $\mathbf{z}_{0j}$ and $\mathbf{u}_j$ can be updated independently in parallel for each $j=1,\hdots,J$.  Following a similar sequence of steps as discussed in Section \ref{sec.II(C)}, the steps in  (\ref{eq.MU-ADMM1}), (\ref{eq.MU-ADMM2}), and (\ref{eq.MU-ADMM3}) are simplified as
\begin{align}
\mathbf{z}_{0j}^{t+1}:=&\argmin_{\mathbf{z}_{0j}\in \tilde{\mathcal{X}}_{\text{mu}}^{JN_{\rm e}} }\frac{\gamma_j}{2}\Vert \mathbf{z}_{0j}\Vert_2^2+\frac{\rho}{2}\Vert \mathbf{z}_{0j}-\mathbf{z}_{0j}^t-\bar{\mathbf{x}}_{0}^t+\mathbf{u}^t+\bar{\mathbf{z}}_0^t\Vert_2^2\label{eq.MU-ADMM4}\\
\bar{\mathbf{x}}_{0}^{t+1}:=&\argmin_{\bar{\mathbf{x}}_{0} } \Vert \mathbf{r}-J\mathbf{H}\bar{\mathbf{x}}_{0}\Vert_2^2+\frac{\rho J}{2}\Vert \bar{\mathbf{x}}_{0}-\bar{\mathbf{x}}_{1}^{t+1}-\mathbf{u}^t\Vert_2^2\label{eq.MU-ADMM5}\\
\mathbf{u}^{t+1}:=&\mathbf{u}^t+\bar{\mathbf{z}}_0^{t+1}-\bar{\mathbf{x}}_0^{t+1} \label{eq.MU-ADMM6}
\end{align}
where $\bar{\mathbf{x}}_0=\frac{1}{J}\sum_{j=1}^J \mathbf{x}_{0j}$ and  $\bar{\mathbf{z}}_0=\frac{1}{J}\sum_{j=1}^J \mathbf{z}_{0j}$. The  solutions obtained by solving (\ref{eq.MU-ADMM4}) and (\ref{eq.MU-ADMM5}) are as follows
\begin{align}
\mathbf{z}_{0j}^{t+1}&=\prod_{[-\alpha_j,\alpha_j]\&[-\beta_j,\beta_j]}\frac{\rho}{(\rho+\gamma_j)}(\mathbf{z}_{0j}^t+\bar{\mathbf{x}}_0^t-\mathbf{u}^t-\bar{\mathbf{z}}_0^t)\label{eq.MU-ADMM8},\forall j=1,\hdots,J.\\
\bar{\mathbf{x}}_0^{t+1}&=(\mathbf{H}^H \mathbf{H} J+\rho \mathbf{I})^{-1}(\mathbf{H}^H \mathbf{r}+\rho(\bar{\mathbf{z}}_0^{t+1}+\mathbf{u}^t)),\label{eq.MU-ADMM9}
\end{align}
where $\prod_{[-\alpha_j,\alpha_j]\&[-\beta_j,\beta_j]}(.)$ denotes the projection of the real part of each entry of the vector onto  $[-\alpha_j,\alpha_j]$ and imaginary part of each entry of the vector  onto $[-\beta_j,\beta_j]$. The step in (\ref{eq.MU-ADMM8}) can be solved in parallel, for  $j=1,\hdots,J$. $\mathbf{I}$ is identity matrix of size $Jd_{\mathrm v}\times Jd_{\mathrm v}$  and $JK\times JK$ for SCMA and DCMA, respectively.
From the definition of $\bar{\mathbf{x}}_0$ given above, the estimated multi-user codeword is given by, $\hat{\mathbf{x}}_{\text{mu}}=J\bar{\mathbf{x}}_0$. Each UE's concatenated sparse codeword $\hat{\mathbf{x}}_{0j}$ can be extracted from the $\hat{\mathbf{x}}_{\text{mu}}$ . Let 
$\tilde{\mathbf{x}}_{0j}$ be the $j$th UE codeword after removing zeros from $\hat{\mathbf{x}}_{0j}$. The minimum Euclidean distance (MED) rule is applied to detect the transmitted codeword index corresponding to each UE,
\begin{align}\label{eq.MU-ADMM7}
\hat{p}_j=\argmin_{ p} \Vert \tilde{\mathbf{x}}_{0j}-\textbf{x}_{j,p}\Vert, \quad  j=1,\hdots,J
\end{align}
where $\textbf{x}_{j,p}$ represents $p$th codeword of  the $j$th UE codebook $\textbf{x}_{j,p}\in \mathcal{X}_j^{K\times M}$. Note that for SCMA, the zeros from $\textbf{x}_{j,p}$ are removed to find $\hat{p}_j$. \textbf{Algorithm }\ref{alg1}  details the steps for detection.

\subsubsection{SMX-CD-NOMA }
The detector of SMX-CD-NOMA needs to detect the signals transmitted from $N_{\rm t}$ antennas of $J$ UEs. Here, the detection is more complex than the SIMO case. The resource-wise processing is adapted to simplify the ADMM-based detection problem. 
Recall   from Section \ref{smx_sub}, that the SMX codeword is given by, $\mathbf{x}_{{\rm{smx}},k}=\sum_{j=1}^{N_{\mathrm u} }\mathbf{x}_{0j,k}$.
The ADMM processing first estimates the resource-wise spatial multiplexed transmitted vector $\hat{\mathbf{x}}_{\text{smx},k}$, for $ k=1,\hdots, K$. Then, the transmitted codeword indices are detected via the MED rule. The ML detection problem is converted into sharing problem as follows
\begin{algorithm}[!htbp]
  \small
 	\caption{SMX CD-NOMA detection via sharing ADMM}
 	\begin{algorithmic}[1]
 	\State  \textbf{Input:} $N_0, E_{\mathrm s}$ and initialize   $\lbrace\mathbf{z}_{0j,k}\rbrace_{j=1}^{N_u}, \bar{\mathbf{x}}_{0,k},\mathbf{u},\bar{\mathbf{z}}_{0,k}$ with zero vectors.  \\	
 	 \textbf{Output:} $\bar{\mathbf{x}}_{0}^{(T)}$.
 	\State From (\ref{eq.SMX-ADMM2}) and (\ref{eq.SMX-ADMM3}),  obtain optimal solutions (\ref{eq.SMX-ADMM6}) and (\ref{eq.SMX-ADMM7}) respectively.
            \State \textbf{Preprocessing}\\
            $\rho=N_0/E_{\mathrm s}$\\
            $\mathbf{G}=(\mathbf{H}_k^H\mathbf{H}_kJ+\rho\mathbf{I})^{-1}$\\
           $\mathbf{M}=\mathbf{H}_k^H\mathbf{r}_k$ 
  
			\For {$t=1,2,,\hdots T $}
 	\State \textbf{Step}:1 Update $\lbrace \mathbf{z}_{0j,k}^{t+1}\rbrace_{j=1}^J$  in parallel  via (\ref{eq.SMX-ADMM6}) and compute $\bar{\mathbf{x}_1}$.
        \State \textbf{Step}:2
        
 	\State Update $\bar{\mathbf{x}}_{0,k}^{t+1}$ via (\ref{eq.SMX-ADMM7}).
	\State Update $\mathbf{u}^{t+1}$ via (\ref{eq.SMX-ADMM3})
		\EndFor
        \State Compute $\hat{\mathbf{x}}_{\text{smx},k}=J\bar{\mathbf{x}}_{0,k}^{(T)}$ (for DCMA), and $\hat{\mathbf{x}}_{\text{smx},k}=d_{\mathrm f}\bar{\mathbf{x}}_{0,k}^{(T)}$ (for SCMA). 
        \State After extracting $\hat{\mathbf{x}}_{j,N_{\mathrm t}}$, apply the MED rule according to (\ref{40}). 	 	 		
 	\end{algorithmic} 
  \label{alg2}
 \end{algorithm}
 \normalfont
\begin{align}\label{eq.SMX-ADMM}
\min_{\mathbf{x}_{0j,k}} \quad &\Vert \mathbf{r}_k-\mathbf{H}_k(\sum_{j=1}^{N_{\mathrm u} }\mathbf{x}_{0j,k})\Vert_2^2+\sum_{j=1}^{N_{\mathrm u}}  \frac{\gamma_j}{2}\Vert \mathbf{x}_{0j,k}\Vert_2^2\\ 
\textrm{s.t.}\quad & \mathbf{x}_{0j,k}\in \tilde{\mathcal{X}}_{j,\text{smx}}^{N_{\mathrm u}N_{\mathrm t}} \nonumber
\end{align} 
where $\gamma_j\geq 0$ is the penalty parameter. By introducing alternative variable $\mathbf{z}_{0j,k}=\mathbf{x}_{0j,k}$,  the ADMM steps for (\ref{eq.SMX-ADMM}) are as follows:
\begin{align}
\mathbf{z}_{0j,k}^{t+1}:=&\argmin_{\mathbf{z}_{0j,k}\in \tilde{\mathcal{X}}_{j,\text{smx}}^{N_{\mathrm u}N_{\mathrm t}} }\frac{\gamma_j}{2}\Vert \mathbf{z}_{0j,k}\Vert_2^2+\frac{\rho}{2}\Vert \mathbf{z}_{0j,k}-\mathbf{x}_{0j,k}^t+\mathbf{u}_{j}^t\Vert_2^2\label{eq.SMX-ADMM1}\\
\mathbf{x}_{0j,k}^{t+1}:=&\argmin_{\mathbf{x}_{0j,k}} \Vert \mathbf{r}_k-\mathbf{H}_k\left(\sum_{j=1}^{N_{\mathrm u}} \mathbf{x}_{0j,k}\right)\Vert_2^2+\frac{\rho}{2}\sum_{j=1}^J\Vert \mathbf{z}_{0j,k}^{t+1}-\mathbf{x}_{0j,k}+\mathbf{u}_j^k\Vert_2^2\label{eq.SMX-ADMM2}\\
\mathbf{u}_j^{t+1}:=&\mathbf{u}_j^k+\left(\mathbf{z}_{0j,k}^{t+1}-\mathbf{x}_{0j,k}^{t+1}\right).\label{eq.SMX-ADMM3}
\end{align}

The  solutions for (\ref{eq.SMX-ADMM1}) and (\ref{eq.SMX-ADMM2}) are as follows:
 \begin{align}
\mathbf{z}_{0j,k}^{t+1}=&\prod_{[-\alpha_j,\alpha_j]\&[-\beta_j,\beta_j]}\frac{\rho}{(\rho+\gamma_j)}(\mathbf{z}_{0j,k}^t+\bar{\mathbf{x}}_{0,k}^t-\mathbf{u}^t-\bar{\mathbf{z}}_{0,k}^t)\label{eq.SMX-ADMM6}\\
\bar{\mathbf{x}}_{0,k}^{t+1}=&(\mathbf{H}_k^H \mathbf{H}_k N_{\mathrm u}+\rho \mathbf{I})^{-1}(\mathbf{H}_k^H \mathbf{r}_k+\rho(\bar{\mathbf{z}}_{0,k}^{t+1}+\mathbf{u}^t))\label{eq.SMX-ADMM7}
\end{align}
where  $\mathbf{I}$ is identity matrix of size $N_{\mathrm u}N_{\mathrm t}\times N_{\mathrm u}N_{\mathrm t}$ and the definitions of $\bar{\mathbf{x}}_{0,k}$ and $ \bar{\mathbf{z}}_{0,k}$ are given below
\begin{align}
\bar{\mathbf{x}}_{0,k}=&\frac{1}{N_{\mathrm u}}\sum_{j=1}^{N_{\mathrm u}} \mathbf{x}_{0j,k} \quad \quad   \bar{\mathbf{z}}_{0,k}=\frac{1}{N_{\mathrm u}}\sum_{j=1}^{N_{\mathrm u}} \mathbf{z}_{0j,k}.\label{eq.SMX-ADMM4}
\end{align}

From the definition of $\bar{\mathbf{x}}_{0,k}$, the estimated codeword corresponding to the $k$th RE is given by $\hat{\mathbf{x}}_{\text{smx},k}=N_{\mathrm u}\bar{\mathbf{x}}_{0,k}$. The sparse vector $\hat{\mathbf{x}}_{0j,k}$  corresponding to  $j$th UE  can be extracted from $ \hat{\mathbf{x}}_{\text{smx},k}$. Let $ \tilde{\mathbf{x}}_{j,k}$ be a  $N_{\mathrm t}\times 1$ vector formed after removing the  zeros from sparse vector $\hat{\mathbf{x}}_{0j,k}$.  Let $\hat{\mathbf{x}}_{ j}^{n_{\mathrm t}}$ be the estimated codeword of $j$th UE corresponding to $n_{\mathrm t}$th transmit antenna. $\hat{\mathbf{x}}_{ j}^{n_{\mathrm t}}$ is formed after resource-wise processing is finished. MED rule is applied to detect the  transmitted codeword index corresponding to each UE and transmit antenna as follows:
\begin{align}\label{40}
\hat{p}_j^{n_{\mathrm t}}=\argmin_{p }  \Vert \hat{\mathbf{x}}_{ \mathrm j}^{n_{\mathrm t}}-\mathbf{x}_{j,p}\Vert, \quad  j=1,\hdots,J,\hspace{2mm} n_{\mathrm t}=1,\hdots,N_{\mathrm t}.
\end{align}
where $\mathbf{x}_{j,p}$ represents the $p$th codeword of  the $j$th UE codebook $\mathcal{X}_j^{K\times M}$. The detailed ADMM-based detection procedure is given in \textbf{Algorithm} \ref{alg2}.
\subsubsection{SM-CD-NOMA}
This section discusses the formulation of the ADMM-based detection problem for the SM-CD-NOMA system. The modulation happens in both the signal domain and the spatial domain. The SM-CD-NOMA system transmits information on the codeword and active antenna indices simultaneously.  The receiver needs to estimate these two quantities. The variable $\mathbf{x}_{\text{sm},k}$ from Section \ref{SIMO_sub1}, is given by $\mathbf{x}_{\text{sm},k}= \sum_{j=1}^{N_{\mathrm u} }\mathbf{x}_{0j,k}^{n_{\rm a}^j}$.
The proposed method estimates the resource-wise spatial modulated transmitted vector $\hat{\mathbf{x}}_{\text{sm},k}$ for $ k=1,\hdots, K$ via ADMM processing. Then, the $L$1-norm rule is applied to detect the antenna index, and the MED rule is applied to detect the transmitted codeword index. The sharing problem of the SM-CD-NOMA system is similar to the problem in (\ref{eq.SMX-ADMM}).
Following the steps (\ref{eq.SMX-ADMM1}), (\ref{eq.SMX-ADMM2}), and (\ref{eq.SMX-ADMM3}), the sharing ADMM problem of SM-CD-NOMA can be solved.  
 The  solutions obtained for the sharing ADMM  problem are similar to (\ref{eq.SMX-ADMM6}) and (\ref{eq.SMX-ADMM7}) and are repeated here for clarity:
 \begin{align}
\mathbf{z}_{0j,\text{sm},k}^{t+1}=&\prod_{[-\alpha_j,\alpha_j]\&[-\beta_j,\beta_j]}\frac{\rho}{(\rho+\gamma_j)}(\mathbf{z}_{0j,\text{sm},k}^t+\bar{\mathbf{x}}_{0,\text{sm},k}^t-\mathbf{u}^t-\bar{\mathbf{z}}_{0,\text{sm},k}^t)\label{eq.SM-ADMM1}\\
\bar{\mathbf{x}}_{0,\text{sm},k}^{t+1}=&(\mathbf{H}_k^H \mathbf{H}_k N_{\mathrm u}+\rho \mathbf{I})^{-1}(\mathbf{H}_k^H \mathbf{r}_k+\rho(\bar{\mathbf{z}}_{0,\text{sm},k}^{t+1}+\mathbf{u}^t)).\label{eq.SM-ADMM2}
\end{align}
The definitions of $\bar{\mathbf{x}}_{0,\text{sm},k}$ and $\bar{\mathbf{z}}_{0,\text{sm},k}$ are similar to (\ref{eq.SMX-ADMM4}). The estimated SM codeword corresponding to $k$th RE is $\hat{\mathbf{x}}_{\text{sm},k}=N_{\mathrm u}\bar{\mathbf{x}}_{0,\text{sm},k}$.  
After resource-wise processing, the structure of the transmitted codewords for $J$ UEs can be retrieved. For\textbf{ Example} \ref{ex3}, the detected codewords for $6$ UEs is given by
\begin{align*}\hat{\mathbf{x}}_{\text{tx}} =
			\begin{bmatrix}
					\hat{\mathbf{x}}_{1}^1&\hat{\mathbf{x}}_{2}^1&	\hat{\mathbf{x}}_{3}^1&\hat{\mathbf{x}}_{4}^1&\hat{\mathbf{x}}_{5}^1&\hat{\mathbf{x}}_{6}^1\\
                 \hat{\mathbf{x}}_{1}^2& \hat{\mathbf{x}}_{2}^2&\hat{\mathbf{x}}_{3}^2&\hat{\mathbf{x}}_{4}^2&\hat{\mathbf{x}}_{5}^2&\hat{\mathbf{x}}_{6}^2\\
			\end{bmatrix}_{N_{\mathrm t}K\times J (8\times 6)}     
		\end{align*} 
where $\hat{\mathbf{x}}_{ j}^{ n_{\mathrm t}}$ indicates the estimated codeword of $j$th UE over the  $n_{\mathrm t}$th transmit antenna. In the presence of noise,  the detected active transmit antenna index is given by
\begin{align}
\hat{n}_{\rm a}^j=\argmax_{n_{\mathrm t}} (\Vert \hat{\mathbf{x}}_{  j}^{\mathrm n_{\mathrm t}}\Vert_1), 
\quad\forall j
\end{align} 
where $\Vert(\cdot)\Vert_1$ denotes the $L$1-norm of $(\cdot)$.
Let $\hat{\mathbf{x}}_{j}^{\hat{n}_{\rm a}^j}$ be the estimated codeword of the $j$th UE transmitted from $\hat{n}_{\rm a}^j$ active antenna. The MED rule is applied to detect the codeword index corresponding to each UE
\begin{align}
\hat{p}_j=\argmin_{ p}  \Vert \hat{\mathbf{x}}_{j}^{\hat{n}_a^j}-\mathbf{x}_{j,p}\Vert, \quad  \forall j.
\end{align}
Similar steps of \textbf{Algorithm} \ref{alg2} are followed in the detection of the SM-CD-NOMA system.

\vspace{-6mm}
 
 \section{Computational Complexity }\label{sec4}
This section analyses the computational complexity of the SIMO-CD-NOMA system. 
The detection algorithm's computational complexity determines its practical viability. The number of FLOPs (mainly complex multiplications) is a useful metric to analyze the complexity of the detector \cite{Wei}. The number of calculations in the proposed CD-NOMA detection via ADMM consists of two parts. Part-1 is iteration-independent (Pre-processing) steps, described in lines 6 and 7 in \textbf{Algorithm}~\ref{alg1}, and Part-2 is  iteration-dependent steps, from lines 9 to 12. The calculations in Part-1 are performed only once, i.e., before the ADMM iterations. In the SIMO-CD-NOMA system, Part-1 contains three steps of calculations such as $\mathbf{H}^H \mathbf{H}$, $(\mathbf{H}^H \mathbf{H}+\rho \mathbf{I})^{-1}$, and $\mathbf{H}^H \mathbf{r}$. The size of $\mathbf{H}$ is $N_{\mathrm r}K\times Jd_{\mathrm v}$ and $N_{\mathrm r}K\times JK$ for SCMA and DCMA, respectively. Further, the numbers of FLOPs required to perform these three steps over the SCMA system are $(N_{\mathrm r} K)(Jd_{\mathrm v})^2,(Jd_{\mathrm v})^3$ and $(N_{\mathrm r} K)(J d_{\mathrm v})$. The numbers of FLOPs required to perform the same steps over the DCMA system are $(N_{\mathrm r} K)(JK)^2,(JK)^3$, and $(N_{\mathrm r} K)(J K)$. The calculations in Part-2 need to be repeated in every iteration and contain mainly two steps.
For SCMA system, these steps involve scalar multiplication of  $Jd_{\mathrm v}\times 1$ vector for $J$ UEs parallelly in (\ref{eq.MU-ADMM8}), multiplication of $Jd_{\mathrm v}\times Jd_{\mathrm v}$ matrix with $Jd_{\mathrm v}\times 1$ and  scalar multiplication  with $Jd_{\mathrm v}\times 1$ vector in (\ref{eq.MU-ADMM9}), in total $J^2d_{\mathrm v}+(Jd_{\mathrm v})^2+Jd_{\mathrm v}$ FLOPs approximately. For DCMA system, $(J^2K+(JK)^2+JK)$ FLOPs are required. The approximate total computational cost to implement the ADMM-based detector over the SIMO-SCMA system and SIMO-DCMA system are  $(N_{\mathrm r} K)(Jd_{\mathrm v})^2+(Jd_{\mathrm v})^3+ (N_{\mathrm r} K)(J d_{\mathrm v})+T(J^2d_{\mathrm v}+(Jd_{\mathrm v})^2+Jd_{\mathrm v})$ and $ (N_{\mathrm r} K)(JK)^2+(JK)^3+(N_{\mathrm r} K)(J K)+T(J^2K+(JK)^2+JK)$ respectively.
 TABLE \ref{T1} compares the total complexity of the sharing ADMM-based detection problem with other known detection schemes. In TABLE \ref{T1}, NA stands for \lq Not  applicable\rq.
 
MPA is widely used for SCMA systems, and its complexity is exponential with $M$ and $d_{\mathrm f}$, as mentioned in TABLE \ref{T1} \cite{Wei}. The computational burden will grow further for large-scale SCMA systems. Further, due to exponential complexity, MPA is not feasible in large-scale MIMO systems. Additionally, the lack of sparsity in DCMA makes MPA impractical for detection. 

A single tree search (STS) based soft-in soft-out (SISO) GSD \cite{Studer} is applied to detect the signals of a spreading sequence-based DCMA system, also known as an overloaded CDMA system~\cite{Liu}. GSD performs key pre-processing steps to convert the rank deficient system into full rank one \cite{Cui}. These steps include  $\mathbf{H}^{H}\mathbf{H}$ in $(\mathbf{Q}=\mathbf{H}^H\mathbf{H}+\lambda \mathbf{I}_{J})$, Cholesky decomposition $\mathbf{Q}=\mathbf{D}^{H}\mathbf{D}$, and $(\mathbf{H}\mathbf{D}^{-1})^H \mathbf{r}$. These steps require $(N_{\mathrm r }K)J^2$, $J^3/3$, and $(J^3+(N_{\mathrm r}K)J^2+JN_{\mathrm r}K)$ FLOPs, respectively. The rank-deficient linear system is thus converted into a full-rank one,  and standard sphere decoding (SD) can be readily applied. The SD performs pre-processing steps, including  $\textbf{QR}$ decomposition and $\mathbf{Q}^H \mathbf{r}$. Further, the numbers of FLOPs to perform these steps are   $J^3$ and $2J^2$, respectively. The major complexity of the SD lies in the tree search algorithm \cite{Hassibi}. The expected complexity of SD is  $E_{\text{FLOPS}}=\sum_{j=1}^J f_p(j)N_j$, where $N_j$ is the average number of nodes visited in level-$j$  of the tree and $f_p(j) $ is the number of FLOPs needed in level-$j$. It is given in  \cite{Hassibi}, $f_p(j)=2j+11$, and $N_j$ is roughly cubic in the number  $J$ of unknowns to be solved.   Fig. \ref{hist} depicts the computational complexity comparison of various detection schemes for different CD-NOMA systems with $N_{\mathrm r}=4, K=4, J=6,d_{\mathrm v}=2$, and $M=4$. It can be observed that the complexity of ADMM-based detector is almost equal to that of the low-complexity MMSE detector. 

\begin{table}
\scriptsize
\centering
\caption{Computational complexity of different detectors for SIMO CD-NOMA systems}
\addtolength{\tabcolsep}{-4pt}
\begin{tabular}{ |p{1.4cm}|c|p{4.4cm}|p{5cm}|p{4.3cm}| } 
 \hline
Detector&\multicolumn{2}{|c|}{Spreading sequence based }  & \multicolumn{2}{|c|}{Codebook based  } \\ \hline
&LDS& \qquad DCMA&  \qquad SCMA& \qquad DCMA \\\hline \hline
 MPA & NA & \qquad NA&$(Kd_{\mathrm f}^2M^{d_{\mathrm f}}N_{\mathrm r}+Nd_{\mathrm f}Md_{\mathrm v})T$ &  \qquad NA \\\hline 
GSD& NA &$(N_{\mathrm r}K)J^2+J^3/3+(J^3+(N_{\mathrm r}K)J^2+JN_{\mathrm r}K)$ + $J^3+2J^2+ E_{\text{FLOPS}}$& \qquad NA & \qquad NA \\  \hline
 MMSE&NA & $(N_{\mathrm r}K)J^2+J^3+(N_{\mathrm r} K) J$ &$(N_{\mathrm r} K)(Jd_{\mathrm v})^2+(Jd_{\mathrm v})^3+(N_{\mathrm r} K)(J d_{\mathrm v})$ & $ (N_{\mathrm r} K)(JK)^2+(JK)^3+(N_{\mathrm r} K)(J K)$\\\hline
 Sharing ADMM &NA &$(N_{\mathrm r}K)J^2+J^3$+ $(N_{\mathrm r} K)J +T(2J^2+J)$ & $N_{\mathrm r} K(Jd_{\mathrm v})^2+(Jd_{\mathrm v})^3$+ $ (N_{\mathrm r} K)(J d_{\mathrm v})+T(J^2d_{\mathrm v}+(Jd_{\mathrm v})^2+Jd_{\mathrm v})$ &$ N_{\mathrm r}K(JK)^2+(JK)^3+(N_{\mathrm r} K)(J K)$ $+T(J^2K+(JK)^2+JK)$ \\ 
 \hline 
\end{tabular}
\label{T1}
\end{table}

\pgfplotsset{every axis/.append style={
line width=0.7 pt, tick style={line width=0.7pt}}, width=8cm,height=5cm, 
 legend style={font=\scriptsize},
 legend pos= north west}
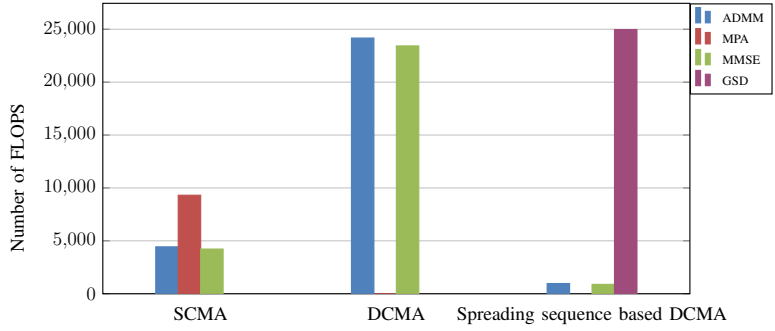
\begin{figure}
    \centering    
   \begin{tikzpicture}[scale=0.6]
    \begin{axis}[
        width  = 0.85*\textwidth,
        height = 8cm,
        major x tick style = transparent,
        ylabel style={at={(-0.05,0.4)}},
        ybar=0pt,
        bar width=14pt,
        ymajorgrids = true,
        ylabel = {Number of FLOPS},
        symbolic x coords={SCMA,DCMA, Spreading sequence based DCMA},
        xtick = data,
        scaled y ticks = false,
        enlarge x limits=0.25,
        ymin=0,
        legend cell align=left,
        legend style={
                at={(1,1)},
                anchor=north west,
                column sep=1ex
        }
    ]
        \addplot[style={bblue,fill=bblue,mark=none}]
            coordinates {(SCMA, 4432) (DCMA,24168) (Spreading sequence based DCMA,966)};

        \addplot[style={rred,fill=rred,mark=none}]
             coordinates {(SCMA,  9312) (DCMA,0)  };

        \addplot[style={ggreen,fill=ggreen,mark=none}]
             coordinates {(SCMA, 4224) (DCMA, 23424) (Spreading sequence based DCMA, 888)};

        \addplot[style={ppurple,fill=ppurple,mark=none}]
             coordinates { (Spreading sequence based DCMA, 24954)};

        \legend{ADMM,MPA,MMSE,GSD}
    \end{axis}  
  \end{tikzpicture}
 \caption{\footnotesize{Computational complexity comparison of various detectors. }}
    \label{hist}
\end{figure}


 \section{Simulations and Discussions}\label{sec.5}

This section presents wide-ranging simulation results of various MIMO-CD-NOMA systems, including  
 SER performance and selection of ADMM parameters $\left(T,\{\gamma\}_{j=1}^J\right)$.    The simulations are performed for various  MIMO-CD-NOMA systems described in Section ~\ref{sec.2} with varying parameters  $\lambda, M, N_{\mathrm r}$, and $N_{\mathrm t}$. Further, the performance of the ADMM-based detector is compared with that of the MPA detector with ten iterations ($T=10$) in the case of the SCMA system. 
The  DCMA and overloaded CDMA systems' detection is carried out using the  MMSE and the GSD detectors, respectively. 
 The codebooks designed in \cite{Vikas} and \cite{Liu} are considered, for SCMA  and DCMA, respectively. Table \ref{table:eva_ch} further details the simulation parameters.  The sample simulation codes for this work  are available at https://github.com/vikas2020-del/ADMM-based-detector-for-NOMA.
\begin{table}[!htpb]
\centering
	\begin{tabular}{|c|c|} \hline
		Parameter & Value \\ \hline \hline
		Modulation order ($M$) & $4,8,16$ \\ \hline
		Number of UEs ($J$) & $6,10$\\  \hline 
            Number of resources ($K$) & $4,5$\\ \hline
		Number of transmit antenna ($N_{\mathrm t}$) & $ 2, 4, 8$\\ \hline 
	Number of receive antenna ($N_{\mathrm r}$) & $ 4, 8, 32, 64$\\  \hline 	Overloading factor ($\lambda$)  & $150~\%, 200~\%$\\  \hline
 Detectors   & ADMM, MPA, MMSE, GSD\\  \hline
 Number of iterations ($T$) & 30 (ADMM), 10 (MPA)\\ \hline
	\end{tabular}
\caption{Simulation parameters.}
\label{table:eva_ch}
\end{table}
\vspace{-4mm}
\subsection{SER performance of SIMO CD-NOMA system}

Fig.~\ref{Fig.2_1(a)} and Fig.~\ref{Fig.2_1(b)} show the SER performance  of SIMO-SCMA system for $150~\%$ and $200~\%$ overloading, respectively.
\pgfplotsset{every semilogy axis/.append style={
line width=0.7 pt, tick style={line width=0.7pt}}, width=8cm,height=7cm, 
legend style={font=\tiny},
legend pos= south west}
\begin{figure}[htb!]
\centering
\begin{subfigure}[b]{0.42\textwidth}
        \centering
        \resizebox{\linewidth}{!}{
\begin{tikzpicture}[new spy style]
\begin{semilogyaxis}[xmin=-5, xmax=5, ymin=1e-06, ymax=1,
xlabel={ $E_b/N_0$ (dB)},
ylabel={SER},
grid=both,
grid style={dotted},
legend cell align=left,
legend entries={ {$N_{\mathrm r}$=16, $M$=8},{$N_{\mathrm r}$=4, $M$=4},{$N_{\mathrm r}$=4, $M$=8},{$N_{\mathrm r}$=8, $M$=8},{$N_{\mathrm r}$=8, $M$=4},{$N_{\mathrm r}$=4, $M$=4},{$N_{\mathrm r}$=8, $M$=4}},
cycle list name=black white
]
\addplot [black,mark=diamond , mark size=3 pt]table [x={x}, y={y1}]
{data/MU_MIMO_150_new.txt};
\addplot [magenta,mark=oplus , mark size=3 pt]table [x={x}, y={y2}]
{data/MU_MIMO_150_new.txt};
\addplot [black,mark=o , mark size=3 pt]table [x={x}, y={y3}]
{data/MU_MIMO_150_new.txt};
\addplot [black,mark=triangle , mark size=3 pt]table [x={x}, y={y4}]
{data/MU_MIMO_150_new.txt};

\addplot [magenta,mark=x , mark size=3 pt]table [x={x}, y={y5}]
{data/MU_MIMO_150_new.txt};

\addplot [magenta, mark=oplus,very thick,dotted, mark size=3 pt,mark options={scale=1,solid}]table [x={x},y={y_10_MMSE1}]
{data/MU_MIMO_150_new.txt};

\addplot [magenta, mark=x,very thick, dotted, mark size=3 pt,mark options={scale=1,solid}]table [x={x},y={y_11_MMSE2}]
{data/MU_MIMO_150_new.txt};


\end{semilogyaxis}
\draw[black] (5.4,5) node[]{$\lambda=150~\%$};
\end{tikzpicture}
}
\caption{\footnotesize{SER vs. SNR performance for  $\lambda=150~\%$.}}
\label{Fig.2_1(a)}
\end{subfigure}
\pgfplotsset{every semilogy axis/.append style={
line width=0.7 pt, tick style={line width=0.7pt}}, width=8cm,height=7cm, 
legend style={font=\tiny},
legend pos= south west}
\begin{subfigure}[b]{0.42\textwidth}
        \centering
        \resizebox{\linewidth}{!}{
        \begin{tikzpicture}[new spy style]
        \begin{semilogyaxis}[xmin=-5, xmax=15, ymin=1e-07, ymax=1,
xlabel={$E_b/N_0$ (dB)},
ylabel={SER},
grid=both,
grid style={dotted},
legend cell align=left,
legend entries={ {$N_{\mathrm r}=4, M=4$},{$N_{\mathrm r}=8,M=4$},{$N_{\mathrm r}=8, M=8$},{$N_{\mathrm r}=4, M=8$},{$N_{\mathrm r}=16, M=8$},{$N_{\mathrm r}=4, M=4$},{$N_{\mathrm r}=8, M=4$}},
cycle list name=black white
]
\addplot [magenta,mark=oplus , mark size=3 pt]table [x={x}, y={y1}]
{data/MU_MIMO_200.txt};
\addplot [magenta,mark=x , mark size=3 pt]table [x={x}, y={y2}]
{data/MU_MIMO_200.txt};
\addplot [black,mark=triangle, mark size=3 pt]table [x={x}, y={y3}]
{data/MU_MIMO_200.txt};
\addplot [black,mark=o, mark size=3 pt]table [x={x}, y={y6}]{data/MU_MIMO_200.txt};
\addplot [black,mark=diamond, mark size=3 pt]table [x={x}, y={y5}]
{data/MU_MIMO_200.txt};

\addplot [magenta, mark=oplus,very thick, dotted, mark size=3 pt,mark options={scale=1,solid}]table [x={x},y={y6_MMSE1}]
{data/MU_MIMO_200.txt};

\addplot [magenta, mark=x,very thick, dotted, mark size=3 pt,mark options={scale=1,solid}]table [x={x},y={y7_MMSE2}]
{data/MU_MIMO_200.txt};

\end{semilogyaxis}
\draw[black] (5.4,5) node[]{$\lambda=200~\%$};
        \end{tikzpicture}
        }
\caption{ \footnotesize{SER vs. SNR performance for  $\lambda=200~\%$. }}
\label{Fig.2_1(b)}
\end{subfigure}
\caption{\footnotesize{SER performance SIMO SCMA system. Solid and dotted lines denote the ADMM and  MMSE performance, respectively.}}
\label{fig2_1}
\end{figure}
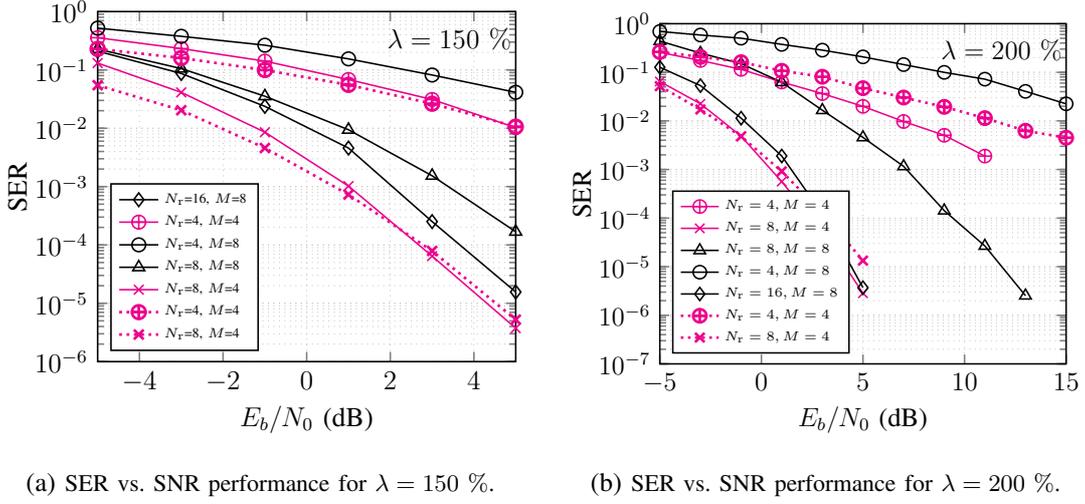
Each curve represents the SER performance for a specific number  $N_{\mathrm r}$  of the receive antennas at the BS and codebook size $M$.
Fig.~\ref{Fig.2_1(a)} and  Fig.~\ref{Fig.2_1(b)} show that  the SER performance improves as $N_{\mathrm r}$ increases. Thus, the proposed ADMM-based detector is able to exploit the diversity gain provided by the SIMO-CD-NOMA system. 
However, for $M=8$, the SER performance is slightly worse than $M=4$ due to the inevitable decrease in the  minimum Euclidean and minimum product distance \cite{Vikas}. 
Fig.~\ref{Fig.2_1(a)} and Fig.~\ref{Fig.2_1(b)} also compare the SER performance of the MMSE and ADMM-based detectors.
 For a small-scale SCMA system ($M=4$, $\lambda=150~\%$), the performance of the MMSE detector is close to that of the ADMM-based detector.
 However, for $\lambda=200~\%$, the MMSE performance is inferior to ADMM. Observe from (\ref{eq1}) that as the number $J$ of UEs  increases, the number $Jd_v$ of unknowns  approaches the number $KN_r$ of observations.  The MMSE performance degrades when the $KN_r $ to $Jd_v$ ratio approaches unity. ADMM provides approximately 2~dB SNR gain at $\text{SER}=10^{-3}$ over the MMSE detector for  200~\% overloaded SCMA system, as observed from Fig.~\ref{Fig.2_1(b)}.
Fig.~\ref{Fig.3_1(a)} and Fig.~\ref{Fig.3_1(b)} show the SER performance  of $M=4$ and $M=16$ overloaded SIMO-DCMA systems, respectively.  Fig.~\ref{Fig.3_1(a)}  compares the SER performance of the MMSE and ADMM-based detectors. ADMM provides approximately 2~dB SNR gain at $\text{SER}=10^{-3}$ over the MMSE detector for the 200~\% overloaded DCMA system, as observed from Fig.~\ref{Fig.3_1(a)}. Therefore, ADMM is a more suitable detector compared with MMSE for 200~\% overloaded CD-NOMA systems. Furthermore,  simulation attempts indicate that MMSE cannot detect the CD-NOMA signals for $M = 8$ and $M = 16$ size codebooks. Therefore, these plots are not shown in Fig. \ref{fig2_1} and Fig. \ref{fig3_1}.

\pgfplotsset{every semilogy axis/.append style={
line width=0.7 pt, tick style={line width=0.7pt}}, width=8cm,height=7cm, 
legend style={font=\scriptsize},
legend pos= north east}
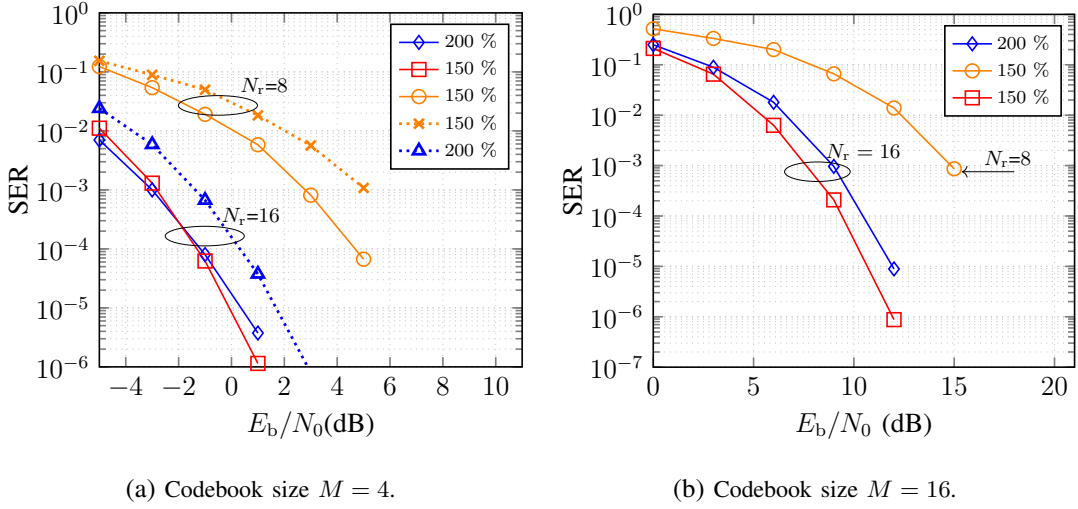
\begin{figure}[htb!]
\centering
\begin{subfigure}[b]{0.42\textwidth}
        \centering
        \resizebox{\linewidth}{!}{
\begin{tikzpicture}[new spy style]
\begin{semilogyaxis}[xmin=-5, xmax=11, ymin=1e-06, ymax=1,
xlabel={$E_{\mathrm b}/N_0$(dB)},
ylabel={SER},
grid=both,
grid style={dotted},
legend cell align=left,
legend entries={ {$200~\%$},{$150~\%$},{$150~\%$},{$150~\%$},{$200~\%$}},
cycle list name=black white
]
\addplot [blue,mark=diamond , mark size=3 pt]table [x={x}, y={y1}]
{data/DCMA_M=4.txt};
\addplot [red,mark=square , mark size=3 pt]table [x={x}, y={y2}]
{data/DCMA_M=4.txt};
\addplot [orange,mark=o , mark size=3 pt]table [x={x}, y={y1}]
{data/DCMA_M=4_1.txt};

\addplot [orange, mark=x,very thick, dotted, mark size=3 pt,mark options={scale=1,solid}]table [x={x}, y={MMSE2}]
{data/DCMA_M=4_1.txt};
\addplot [blue, mark=triangle,very thick, dotted, mark size=3 pt,mark options={scale=1,solid}]table [x={x}, y={MMSE3}]
{data/DCMA_M=4_1.txt};

\end{semilogyaxis}
\draw[black] (1.8,4) ellipse (.6 and 0.15)node[xshift=7mm, yshift=3mm]{\footnotesize{$N_{\mathrm r}$=$8$}};
\draw[black] (1.6,2) ellipse (.6 and 0.15)node[xshift=7mm, yshift=3mm]{\footnotesize{$N_{\mathrm r}$=$16$}};
\end{tikzpicture}
}
\caption{ \footnotesize{Codebook size $M=4$}. }
\label{Fig.3_1(a)}
\end{subfigure}
\begin{subfigure}[b]{0.42\textwidth}
        \centering
        \resizebox{\linewidth}{!}{
\begin{tikzpicture}[new spy style]
\begin{semilogyaxis}[xmin=0, xmax=21, ymin=1e-07, ymax=1,
xlabel={$E_{\mathrm b}/N_0$ (dB)},
ylabel={SER},
grid=both,
grid style={dotted},
legend cell align=left,
legend entries={ {$200~\%$},{$150~\%$},{$150~\%$}},
cycle list name=black white
]
\addplot [blue,mark=diamond , mark size=3 pt]table [x={x}, y={y1}]
{data/DCMA_M=16.txt};
\addplot [orange,mark=o , mark size=3 pt]table [x={x}, y={y2}]
{data/DCMA_M=16.txt};
\addplot [red,mark=square , mark size=3 pt]table [x={x}, y={y3}]
{data/DCMA_M=16.txt};
%

\end{semilogyaxis}
\draw[black] (2.5,3) ellipse (.5 and 0.15)node[xshift=7mm, yshift=3mm]{\footnotesize{$N_{\mathrm r}=16$}};
\draw[black, ->] (5.5,3) --(4.7,3)
node[xshift=7mm, yshift=2mm]{\footnotesize{$N_{\mathrm r}$=$8$}};
\end{tikzpicture}
}
\caption{ \footnotesize{Codebook size $M=16$}. }
\label{Fig.3_1(b)}
\end{subfigure}
\caption{\footnotesize{SER performance of SIMO DCMA system for 150~\% and 200~\% overloading. Solid and dotted lines denote the ADMM and  MMSE performance, respectively. } }
\label{fig3_1}
\end{figure}
\FloatBarrier
\vspace{-7mm}

\subsection{SER performance of SMX CD-NOMA systems  }

Fig.~\ref{Fig. 4(a)} and Fig.~\ref{Fig. 4(b)}  show the SER performance  of  SMX-SCMA system for $150~\%$ and $200~\%$ overloading, respectively. 
Observe from Fig.~\ref{Fig. 4(a)} and Fig.~\ref{Fig. 4(b)}, that the higher the value of $N_{\mathrm{r}}$, the better is the SER performance. As the value of $N_{\mathrm{t}}$ increases,  $N_{\mathrm{r}}$ is increased to maintain $N_{\mathrm{r}}>d_{\mathrm f}N_{\mathrm{t}}$ in (\ref{eq2}). The improvement in the SER performance is maximum for $N_{\mathrm{r}}=64$ case due to more observations at the BS. 
This improvement comes at the cost of a  marginal computational complexity increment (TABLE \ref{T1}) in ADMM detection. Fig.~\ref{Fig. 4(c)} depicts the SER performance of SMX-DCMA systems. For SCMA, $N_{\rm u}=d_{\rm f}$ UEs overlap on each RE. On the other hand, for DCMA $N_{\rm u}=J\; (J>d_{\rm f})$ UEs overlap on each RE, as shown in Fig. \ref{dcma_model}. This significant overlapping increases the number of unknowns in a DCMA system  compared to the SCMA one, as observed in (\ref{eq2}). To maintain the number of observations higher than the number of unknowns ($N_{\mathrm{r}}>JN_{\mathrm{t}}$), DCMA requires larger $N_{\mathrm{r}}$ at the BS as compared to SCMA for a fixed $N_{\mathrm{t}}$. As a result, the computational complexity of DCMA detection significantly increases. The tree search-based SD algorithms are highly complex in codebook-based DCMA, as explained in Section~\ref{sec4}. Further, the ADMM-based detector shows similar complexity as that of MMSE, as shown in Fig. \ref{hist}.  Therefore, compared with MMSE, the ADMM-based detector performs well  with almost the same complexity.

\pgfplotsset{every semilogy axis/.append style={
line width=0.7 pt, tick style={line width=0.7pt}}, width=8cm,height=7cm, 
legend style={font=\tiny},
legend pos= south west}
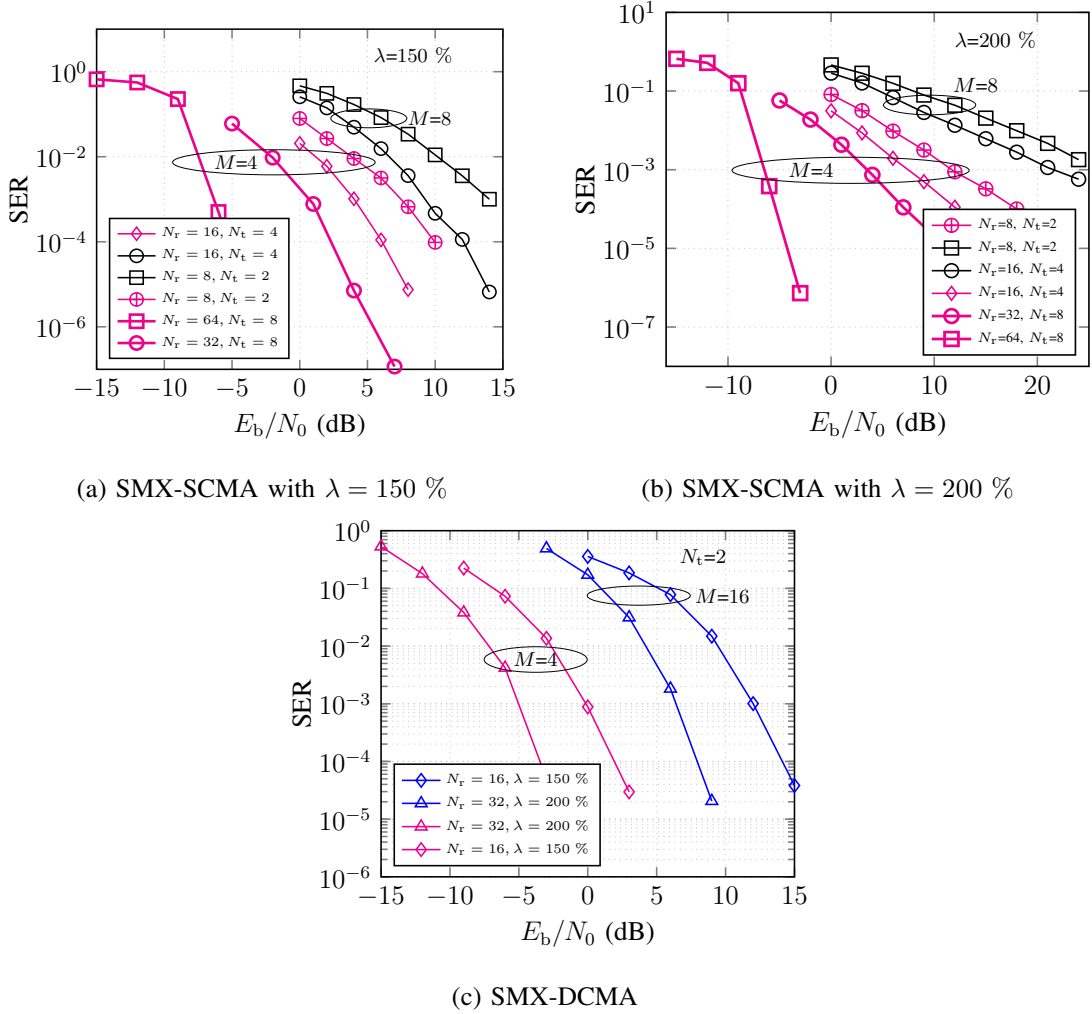
\begin{figure}[htb!]
\centering
\begin{subfigure}[b]{0.42\textwidth}
        \centering
        \resizebox{\linewidth}{!}{
\begin{tikzpicture}[new spy style]
\begin{semilogyaxis}[xmin=-15, xmax=15, ymin=1e-07, ymax=10,
xlabel={$E_{\mathrm b}/N_0$ (dB)},
ylabel={SER},
grid=both,
grid style={dotted},
legend cell align=left,
legend entries={ {$N_{\mathrm r}=16,N_{\mathrm t}=4$},{$N_{\mathrm r}=16,N_{\mathrm t}=4$},{$N_{\mathrm r}=8,N_{\mathrm t}=2$},{$N_{\mathrm r}=8,N_{\mathrm t}=2$},{$N_{\mathrm r}=64,N_{\mathrm t}=8$},{$N_{\mathrm r}=32,N_{\mathrm t}=8$}},
cycle list name=black white
]
\addplot [magenta,mark=diamond , mark size=3 pt]table [x={x}, y={y1}]
{data/SMX_150.txt};
\addplot [black,mark=o , mark size=3 pt]table [x={x}, y={y2}]
{data/SMX_150.txt};
\addplot [black,mark=square , mark size=3 pt]table [x={x}, y={y3}]
{data/SMX_150.txt};
\addplot [magenta,mark=oplus , mark size=3 pt]table [x={x}, y={y4}]
{data/SMX_150.txt};

\addplot [magenta, mark=square, very thick, mark size=3 pt,mark options={scale=1,solid}]table [x={x},y={y_1}]
{data/SMX_150_LM2.txt};

\addplot [magenta, mark=o,very thick,  mark size=3 pt,mark options={scale=1,solid}]table [x={x},y={y_1}]
{data/SMX_150_LM1.txt};

\end{semilogyaxis}
\draw[black] (5,5) node[]{\footnotesize{$\lambda$=$150~\%$}};
\draw[black] (2.8,3.3) ellipse (1.6 and 0.2)node[xshift=-6mm]{\footnotesize{$M$=$4$}};
\draw[black] (4.3,4) ellipse (.6 and 0.15)node[xshift=9.6mm, yshift=0mm]{\footnotesize{$M$=$8$}};
\end{tikzpicture}
}
\caption{ SMX-SCMA with $\lambda=150~\%$ }
\label{Fig. 4(a)}
\end{subfigure}
\pgfplotsset{every semilogy axis/.append style={
line width=0.7 pt, tick style={line width=0.7pt}}, width=8cm,height=7cm, 
legend style={font=\tiny},
legend pos= south east}
\begin{subfigure}[b]{0.42\textwidth}
        \centering
        \resizebox{\linewidth}{!}{
        \begin{tikzpicture}[new spy style]
        \begin{semilogyaxis}[xmin=-16, xmax=25, ymin=1e-08, ymax=10,
xlabel={$E_{\mathrm b}/N_0$ (dB)},
ylabel={SER},
grid=both,
grid style={dotted},
legend cell align=left,
legend entries={ {$N_{\mathrm r}$=8, $N_{\mathrm t}$=2},{$N_{\mathrm r}$=8, $N_{\mathrm t}$=2},{$N_{\mathrm r}$=16, $N_{\mathrm t}$=4},{$N_{\mathrm r}$=16, $N_{\mathrm t}$=4},{$N_{\mathrm r}$=32, $N_{\mathrm t}$=8},{$N_{\mathrm r}$=64, $N_{\mathrm t}$=8}},
cycle list name=black white
]
 \addplot [magenta,mark=oplus , mark size=3 pt]table [x={x}, y={y1}]
 {data/SMX_200.txt};
\addplot [black,mark=square , mark size=3 pt]table [x={x}, y={y2}]
{data/SMX_200.txt};
 \addplot [black,mark=o , mark size=3 pt]table [x={x}, y={y3}]
{data/SMX_200.txt};
\addplot [magenta,mark=diamond , mark size=3 pt]table [x={x}, y={y4}]
 {data/SMX_200.txt};

 \addplot [magenta, mark=o,very thick,  mark size=3 pt,mark options={scale=1,solid}]table [x={x},y={N32}]
 {data/SMX_200_LM.txt};

 \addplot [magenta, mark=square,very thick,  mark size=3 pt,mark options={scale=1,solid}]table [x={x},y={N64}]
 {data/SMX_200_LM.txt};


		\end{semilogyaxis}
  \draw[black] (5,5) node[]{\footnotesize{$\lambda$=$200~\%$}};
\draw[black] (2.8,3) ellipse (1.8 and 0.2)node[xshift=-6mm]{\footnotesize{$M$=$4$}};
\draw[black] (4,4) ellipse (.7 and 0.15)node[xshift=7mm, yshift=3mm]{\footnotesize{$M$=$8$}};
        \end{tikzpicture}
        }
\caption{ SMX-SCMA with $\lambda=200~\%$ }
\label{Fig. 4(b)}
\end{subfigure}

\pgfplotsset{every semilogy axis/.append style={
line width=0.7 pt, tick style={line width=0.7pt}}, width=8cm,height=7cm, 
legend style={font=\tiny},
legend pos= south west}
\begin{subfigure}[t]{0.42\textwidth}
        \centering
        \resizebox{\linewidth}{!}{
\begin{tikzpicture}[new spy style]
\begin{semilogyaxis}[xmin=-15, xmax=15, ymin=1e-06, ymax=1,
xlabel={$E_{\mathrm b}/N_0$ (dB)},
ylabel={SER},
grid=both,
grid style={dotted},
legend cell align=left,
legend entries={ {$N_{\mathrm r}=16, \lambda=150~\%$},{$N_{\mathrm r}=32, \lambda=200~\%$},{$N_{\mathrm r}=32,\lambda=200~\%$},{$N_{\mathrm r}=16, \lambda=150~\%$}},
cycle list name=black white
]
\addplot [blue,mark=diamond , mark size=3 pt]table [x={x}, y={y1(M=16,150)}]{data/DCMA_SMX_new.txt};
 \addplot [blue,mark=triangle , mark size=3 pt]table [x={x}, y={y1(M=16,200)}]{data/DCMA_SMX_new.txt};
\addplot [magenta,mark=triangle , mark size=3 pt]table [x={x}, y={y1(M=4,200)}]{data/DCMA_SMX_new.txt};
   \addplot [magenta,mark=diamond , mark size=3 pt]table [x={x},
  y={y1(M=4,150)}]{data/DCMA_SMX_new.txt};
\end{semilogyaxis}
\draw[black] (5,5) node[]{\footnotesize{$N_{\mathrm t}$=$2$}};
\draw[black] (2.4,3.4) ellipse (.8 and 0.2)node[]{\footnotesize{$M$=$4$}};
\draw[black] (4,4.4) ellipse (.8 and 0.15)node[xshift=13mm]{\footnotesize{$M$=$16$}};
\end{tikzpicture}}
\caption{ SMX-DCMA }
\label{Fig. 4(c)}
\end{subfigure}
\caption{\footnotesize{SER vs. SNR performance of SMX-CD-NOMA system by varying $N_{\mathrm t}, N_{\mathrm r}, M$ using the ADMM-based detector. }  }
\label{fig4}
\end{figure}

\FloatBarrier
\vspace{-0.5mm}
\subsection{SER performance of SM CD-NOMA systems  }
Fig.~\ref{Fig. 5(a)} and Fig.~\ref{Fig. 5(b)}  show the SER performance  of SM-SCMA system for $150~\%$ and $200~\%$ overloading, respectively. The performance of the ADMM-based detector mainly depends on $N_{\mathrm t}, N_{\mathrm r} $, and $M$, as we have seen in the previous subsections.
The $M=4$ size codebook exhibits superior performance over $M=8$ due to its better distance properties (Euclidean distance and product distance). In SM-CD-NOMA systems, each UE's codeword contains the active antenna and codeword index information of that particular UE, as given in (\ref{eq::sm_tx}). Fig.~\ref{Fig. 5(a)} and Fig.~\ref{Fig. 5(b)} depict the efficiency of ADMM in detecting the information both in the signal domain and space domain. Note that the codewords in (\ref{eq::sm_tx}) contain zero-column vectors to indicate zero power transmission from the inactive antenna. 
As a result, the number of observations goes below the number of unknowns ($N_{\mathrm{r}}<N_{\mathrm{u}}N_{\mathrm{t}}$). Thus, it leads to performance loss for an ADMM-based detector. However, increasing the number of observations ($N_{\mathrm r}$) at the BS can compensate for this loss. Therefore, the performance of the ADMM-based detector can be maintained by a marginal increment in the computational complexity as given in  TABLE \ref{T1}.  
The observations mentioned above for SMX-DCMA and SM-SCMA systems apply to SM-DCMA systems, as shown in Fig. \ref{Fig. 5(c)}. Further, $N_{\mathrm{r}}=32, \lambda=200~\%$ provides significantly better performance than $N_{\mathrm{r}}=16, \lambda=150~\%$, as the former has  more observations ($N_{\rm r}$) than the latter. So the ADMM-based detector is more appropriate for highly overloaded DCMA systems equipped with large-scale MIMO parameters. 
\pgfplotsset{every semilogy axis/.append style={
line width=0.7 pt, tick style={line width=0.7pt}}, width=8cm,height=7cm, 
legend style={font=\tiny},
legend pos= south west}
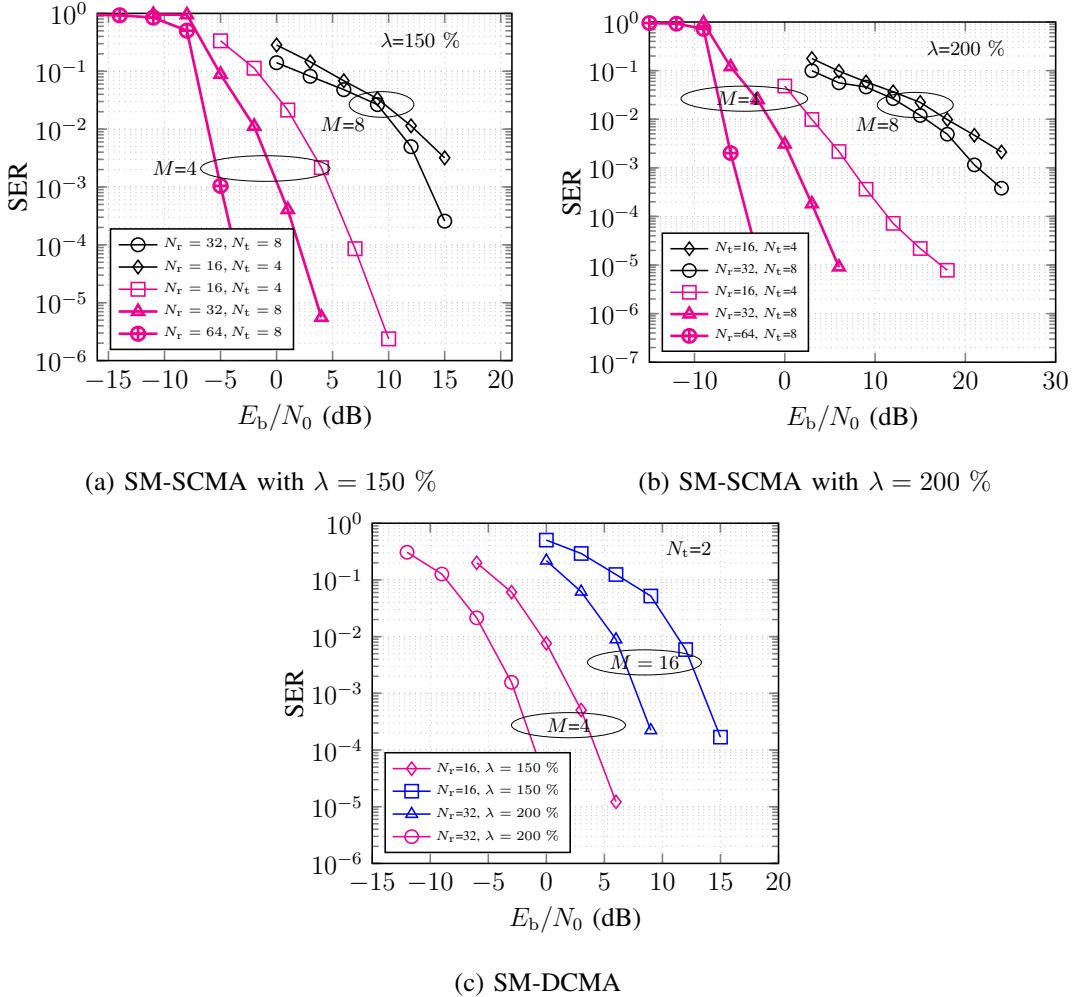
\begin{figure}[htb!]
\centering
\begin{subfigure}[b]{0.42\textwidth}
        \centering
        \resizebox{\linewidth}{!}{
\begin{tikzpicture}[new spy style]
\begin{semilogyaxis}[xmin=-16, xmax=21, ymin=1e-06, ymax=1,
xlabel={ $E_{\mathrm b}/N_0$ (dB)},
ylabel={SER},
grid=both,
grid style={dotted},
legend cell align=left,
legend entries={ {$N_{\mathrm r}=32,N_{\mathrm t}=8$},{$N_{\mathrm r}=16,N_{\mathrm t}=4$},{$N_{\mathrm r}=16,N_{\mathrm t}=4$},{$N_{\mathrm r}=32,N_{\mathrm t}=8$},{$N_{\mathrm r}=64,N_{\mathrm t}=8$}},
cycle list name=black white
]
\addplot [black,mark=o , mark size=3 pt]table [x={x}, y={y1}]
{data/SM_150_new_M=8.txt};
\addplot [black,mark=diamond, mark size=3 pt]table [x={x}, y={y2}]
{data/SM_150_new_M=8.txt};
\addplot [magenta,mark=square , mark size=3 pt]table [x={x}, y={y1}]
{data/SM_150_4.txt};

\addplot [magenta, mark=triangle,very thick,  mark size=3 pt,mark options={scale=1,solid}]table [x={x},y={N32}]
{data/SM_150_LM.txt};

\addplot [magenta, mark=oplus,very thick,  mark size=3 pt,mark options={scale=1,solid}]table [x={x},y={N64}]
{data/SM_150_LM.txt};

\end{semilogyaxis}

 \draw[black] (5,5) node[]{\footnotesize{$\lambda$=$150~\%$}};
\draw[black] (2.6,3) ellipse (1 and 0.2)node[xshift=-14mm]{\footnotesize{$M$=$4$}};
\draw[black] (4.4,4) ellipse (0.5 and 0.2)node[xshift=-6mm, yshift=-3mm]{\footnotesize{$M$=$8$}};
\end{tikzpicture}
}
\caption{SM-SCMA with $\lambda=150~\%$}
\label{Fig. 5(a)}
\end{subfigure}
\begin{subfigure}[b]{0.42\textwidth}
        \centering
        \resizebox{\linewidth}{!}{
        \begin{tikzpicture}[new spy style]
        \begin{semilogyaxis}[xmin=-15, xmax=30, ymin=1e-07, ymax=1,
xlabel={$E_{\mathrm b}/N_0$ (dB)},
ylabel={SER},
grid=both,
grid style={dotted},
legend cell align=left,
legend entries={ {$N_{\mathrm t}$=16, $N_{\mathrm t}$=4}, {$N_{\mathrm r}$=32, $N_{\mathrm t}$=8},{$N_{\mathrm r}$=16, $N_{\mathrm t}$=4},{$N_{\mathrm r}$=32, $N_{\mathrm t}$=8},{$N_{\mathrm r}$=64, $N_{\mathrm t}$=8}},
cycle list name=black white
]
\addplot [black,mark=diamond , mark size=3 pt]table [x={x}, y={y2}]
{data/SM_200_new_M=8.txt};
\addplot [black,mark=o, mark size=3 pt]table [x={x}, y={y3}]
{data/SM_200_new_M=8.txt};
\addplot [magenta,mark=square , mark size=3 pt]table [x={x}, y={y1}]
{data/SM_200.txt};
\addplot [magenta,mark=triangle,very thick, mark size=3 pt]table [x={x}, y={y2}]
{data/SM_200.txt};

\addplot [magenta,mark=oplus ,very thick, mark size=3 pt]table [x={x}, y={y3}]
{data/SM_200.txt};
		\end{semilogyaxis}
  \draw[black] (5,5) node[]{\footnotesize{$\lambda$=$200~\%$}};
  
\draw[black] (1.5,4.2) ellipse (1 and 0.2)node[xshift=-1mm]{\footnotesize{$M$=$4$}};
\draw[black] (4.2,4.1) ellipse (.6 and 0.2)node[xshift=-6mm, yshift=-3mm]{\footnotesize{$M$=$8$}};
        \end{tikzpicture}
        }
\caption{ SM-SCMA with $\lambda=200~\%$ }
\label{Fig. 5(b)}
\end{subfigure}
\pgfplotsset{every semilogy axis/.append style={
line width=0.7 pt, tick style={line width=0.7pt}}, width=8cm,height=7cm, 
legend style={font=\tiny},
legend pos= south west}
\begin{subfigure}[t]{0.42\textwidth}
        \centering
        \resizebox{\linewidth}{!}{
\begin{tikzpicture}[new spy style]
\begin{semilogyaxis}[xmin=-15, xmax=20, ymin=1e-06, ymax=1,
xlabel={$E_{\mathrm b}/N_0$ (dB)},
ylabel={SER},
grid=both,
grid style={dotted},
legend cell align=left,
  legend entries={ {$N_{\mathrm r}$=16, $\lambda=150~\%$},{$N_{\mathrm r}$=16,  $\lambda=150~\%$},{$N_{\mathrm r}$=32,  $\lambda=200~\%$},{$N_{\mathrm r}$=32, $\lambda=200~\%$}},
cycle list name=black white
]
\addplot [magenta,mark=diamond , mark size=3 pt]table [x={x}, y={y1}]
{data/DCMA_SM_150_new.txt};
\addplot [blue,mark=square , mark size=3 pt]table [x={x}, y={y5_150_M16}]
{data/DCMA_SM_150_new.txt};

\addplot [blue,mark=triangle , mark size=3 pt]table [x={x}, y={y4_200_M16}]
  {data/DCMA_SM_150_new.txt};
  \addplot [magenta,mark=o, mark size=3 pt]table [x={x}, y={y4_200_M4}]
  {data/DCMA_SM_150_new.txt};
\end{semilogyaxis}

\draw[black] (3.1,2.2) ellipse (.9 and 0.2)node[]{\footnotesize{$M$=$4$}};
 \draw[black] (4.3,3.2) ellipse (.9 and 0.2)node[]{\footnotesize{$M=16$}};
\draw[black] (5,5) node[]{\footnotesize{$N_{\mathrm t}$=$2$}};
\end{tikzpicture}
}
\caption{  SM-DCMA}
\label{Fig. 5(c)}
\end{subfigure}
\caption{\footnotesize{SER vs. SNR performance for SM-MIMO CD-NOMA system using the proposed ADMM-based detector.}}
\label{fig5}
\end{figure}

\FloatBarrier
\vspace{-5mm}

\subsection{SER performance: ADMM vs. MPA  }
Fig.~\ref{Fig. 7(a)} shows the SER performance comparison between the ADMM and  MPA detectors for the SIMO-SCMA system. Existing research shows that the MPA is a near-optimal detection with high computational complexity \cite{Zhang}. MPA gives a maximum SNR gain of around 2.5~dB over ADMM at $\text{SER}=10^{-2}$ for $N_{\mathrm r}=4$ and $M=4$, as shown in Fig.~\ref{Fig. 7(a)}. As the modulation order $M$ increases, the performance of ADMM becomes close to MPA, as observed for $ N_{\mathrm{r}}=4, M=8$ in Fig.~\ref{Fig. 7(a)}. Therefore, for $M=8$, codebook distance properties  influence MPA and ADMM-based detectors similarly. Further, for $N_{\mathrm{r}}=8, M=8$, the ADMM-based detector 
 shows approximately 2~dB SNR gain at $\text{SER}=10^{-3}$ over the MPA detector. The increase in the number of observations significantly improves the ADMM performance compared to MPA for large-size codebooks (i.e. with large $M$).

\pgfplotsset{every semilogy axis/.append style={
line width=0.7 pt, tick style={line width=0.7pt}}, width=8cm,height=7cm, 
legend style={font=\tiny},
legend pos= south west}
\begin{figure}[htb!]
\centering
\begin{subfigure}[b]{0.42\textwidth}
        \centering
        \resizebox{\linewidth}{!}{
\begin{tikzpicture}[new spy style]
\begin{semilogyaxis}[xmin=-5, xmax=5, ymin=1e-06, ymax=1,
xlabel={$E_{\mathrm b}/N_0$ (dB)},
ylabel={SER},
grid=both,
grid style={dotted},
legend cell align=left,
legend entries={ {$N_{\mathrm r}=4,M=4$, ADMM},{$N_{\mathrm r}=4, M=8$, ADMM},{$N_{\mathrm r}=8, M=8$, ADMM},{$N_{\mathrm r}=8, M=4$, ADMM},{$N_{\mathrm r}=4,M=4$, MPA},{$N_{\mathrm r}=4,M=8$, MPA},{$N_{\mathrm r}=8,M=8$, MPA},{$N_{\mathrm r}=8,M=4$, MPA} },
cycle list name=black white
]
\addplot [magenta,mark=square , mark size=3 pt]table [x={x}, y={y2}]
{data/MU_MIMO_150.txt};
\addplot [blue,mark=o , mark size=3 pt]table [x={x}, y={y3}]
{data/MU_MIMO_150.txt};
\addplot [green,mark=triangle , mark size=3 pt]table [x={x}, y={y4}]
{data/MU_MIMO_150.txt};

\addplot [black,mark=diamond , mark size=3 pt]table [x={x}, y={y5}]
{data/MU_MIMO_150.txt};

\addplot [magenta, mark=square ,very thick, dashed, mark size=3 pt,mark options={scale=1,solid}]table [x={x},y={y6_MPA}]
{data/MU_MIMO_150.txt};

\addplot [blue, mark=o,very thick, dashed, mark size=3 pt,mark options={scale=1,solid}]table [x={x},y={y8_MPA_M8}]
{data/MU_MIMO_150.txt};
\addplot [green, mark=triangle,very thick, dashed, mark size=3 pt,mark options={scale=1,solid}]table [x={x},y={y9_MPA_M8}]
{data/MU_MIMO_150.txt};
\addplot [black, mark=diamond,very thick, dashed, mark size=3 pt,mark options={scale=1,solid}]table [x={x},y={y7_MPA}]
{data/MU_MIMO_150.txt};
\end{semilogyaxis}
\end{tikzpicture}
}
\caption{\footnotesize{SIMO SCMA: ADMM vs. MPA with $\lambda$=$150~\%$}}
\label{Fig. 7(a)}
\end{subfigure}
\pgfplotsset{every semilogy axis/.append style={
line width=0.7 pt, tick style={line width=0.7pt}}, width=8cm,height=7cm, 
legend style={font=\tiny},
legend pos= north east}
\begin{subfigure}[b]{0.42\textwidth}
        \centering
        \resizebox{\linewidth}{!}{
        \begin{tikzpicture}[new spy style]
        \begin{semilogyaxis}[xmin=-5, xmax=5, ymin=1e-07, ymax=1,
xlabel={$E_{\mathrm b}/N_0$ (dB)},
ylabel={SER},
grid=both,
grid style={dotted},
]
\addplot [blue,mark=diamond ,very thick, dashdotted, mark size=3 pt,mark options={scale=1,solid}]table [x={x}, y={y1}]
{data/GSD_Vs_ADMM.txt};
\addplot [magenta,mark=diamond , mark size=3 pt]table [x={x}, y={y2}]
{data/GSD_Vs_ADMM.txt};
\addplot [magenta,mark=square , mark size=3 pt]table [x={x}, y={y3}]
{data/GSD_Vs_ADMM.txt};
\addplot [blue,mark=square,very thick, dashdotted, mark size=3 pt,mark options={scale=1,solid}]table [x={x}, y={y4}]
{data/GSD_Vs_ADMM.txt};
\legend{ {$N_{\mathrm r}=8,M=16$, GSD},{$N_{\mathrm r}=8,M=16$, ADMM},{$N_{\mathrm r}=4,M=4$, ADMM},{$N_{\mathrm r}=4,M=4$, GSD}} 
\end{semilogyaxis}
  \draw[black] (2,1) node[]{\footnotesize{$\lambda$=$150~\%$}};
        \end{tikzpicture}
        }
\caption{\footnotesize{ SIMO Overloaded CDMA: ADMM vs. GSD}}
\label{Fig. 7(b)}
\end{subfigure}
\caption{\footnotesize{Comparison of SER performance using the proposed  ADMM and the conventional  MPA and GSD detectors.}}
\label{fig2}
\end{figure}
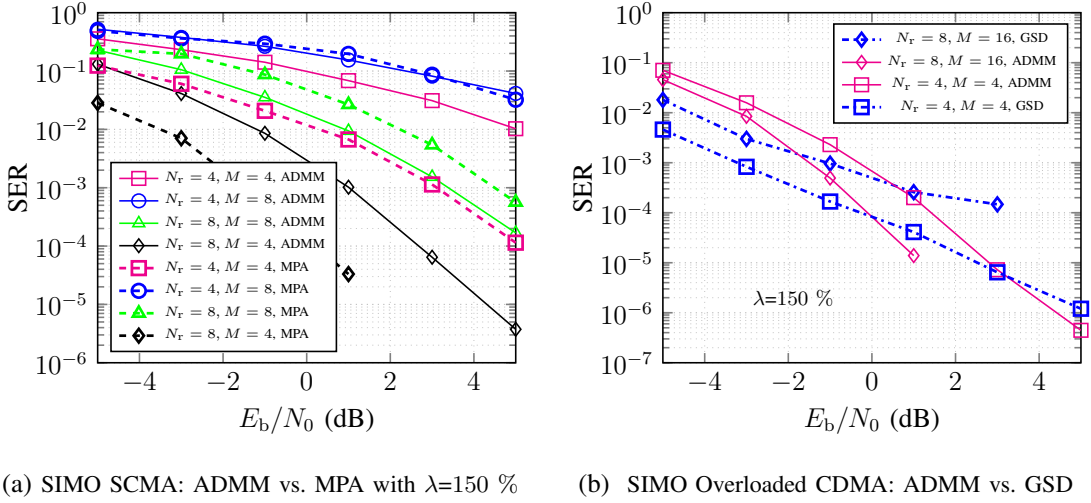
\pgfplotsset{every semilogy axis/.append style={
line width=0.7 pt, tick style={line width=0.7pt}}, width=8cm,height=7cm, 
legend style={font=\scriptsize},
legend pos= north east}
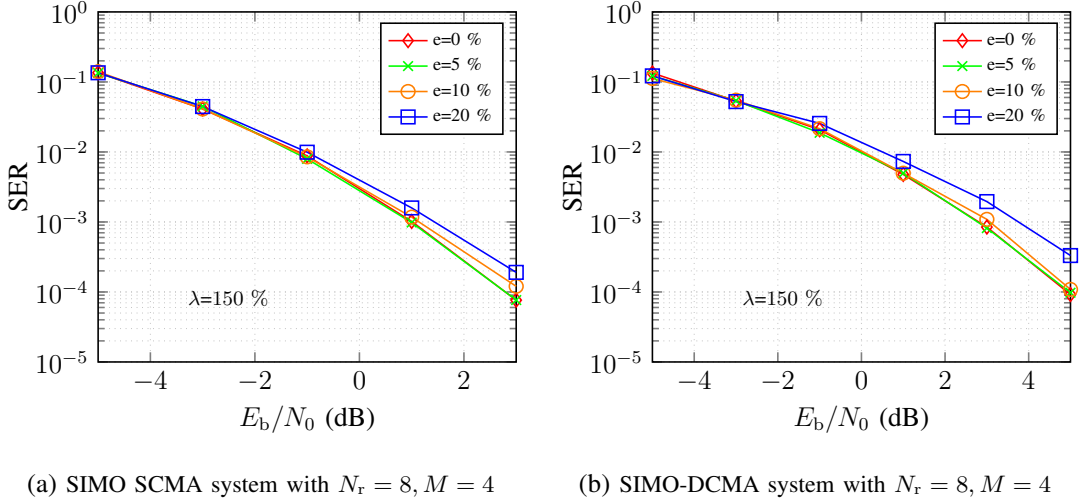
\begin{figure*}[t!]
\centering
\begin{subfigure}[b]{0.42\textwidth}
        \centering
        \resizebox{\linewidth}{!}{
\begin{tikzpicture}[new spy style]
\begin{semilogyaxis}[xmin=-5, xmax=3, ymin=1e-05, ymax=1,
xlabel={$E_{\mathrm b}/N_0$ (dB)},
ylabel={SER},
grid=both,
grid style={dotted},
legend cell align=left,
legend entries={ { e=0~\%},{e=5~\%},{e=10~\%},{e=20~\%}},
cycle list name=black white
]
\addplot [red,mark=diamond , mark size=3 pt]table [x={x}, y={y1}]
{data/CEE/SCMA_CEE.txt};
\addplot [green,mark=x , mark size=3 pt]table [x={x}, y={y4}]
{data/CEE/SCMA_CEE.txt};
\addplot [orange,mark=o , mark size=3 pt]table [x={x}, y={y3}]
{data/CEE/SCMA_CEE.txt};
\addplot [blue,mark=square , mark size=3 pt]table [x={x}, y={y2}]
{data/CEE/SCMA_CEE.txt};

\end{semilogyaxis}
 \draw[black] (2,1) node[]{\footnotesize{$\lambda$=$150~\%$}};
\end{tikzpicture}
}
\caption{ \footnotesize{SIMO SCMA system with $N_{\mathrm r}=8, M=4$}}
\label{Fig. 8(a)}
\end{subfigure}
\begin{subfigure}[b]{0.42\textwidth}
        \centering
        \resizebox{\linewidth}{!}{
\begin{tikzpicture}[new spy style]
\begin{semilogyaxis}[xmin=-5, xmax=5, ymin=1e-05, ymax=1,
xlabel={$E_{\mathrm b}/N_0$ (dB)},
ylabel={SER},
grid=both,
grid style={dotted},
legend cell align=left,
legend entries={{e=0~\%},{e=5~\%}, {e=10~\%},{e=20~\%}},
cycle list name=black white
]
\addplot [red,mark=diamond , mark size=3 pt]table [x={x}, y={y3}]
{data/CEE/DCMA_CEE.txt};
\addplot [green,mark=x , mark size=3 pt]table [x={x}, y={y4}]
{data/CEE/DCMA_CEE.txt};
\addplot [orange,mark=o , mark size=3 pt]table [x={x}, y={y1}]
{data/CEE/DCMA_CEE.txt};
\addplot [blue,mark=square , mark size=3 pt]table [x={x}, y={y2}]
{data/CEE/DCMA_CEE.txt};

%

\end{semilogyaxis}
 \draw[black] (2,1) node[]{\footnotesize{$\lambda$=$150~\%$}};
\end{tikzpicture}
}
\caption{ \footnotesize{SIMO-DCMA system with $N_{\mathrm r}=8, M=4$ }}
\label{Fig. 8(b)}
\end{subfigure}
\caption{\footnotesize{SER performance of CD-NOMA system in imperfect CSI for $e=0~\%, e=5~\%, e=10~\%, e=20~\%$.}} 
\label{Fig.8}
\end{figure*}

\vspace{-5mm}

\subsection{SER performance: ADMM vs. GSD  }
Fig.~\ref{Fig. 7(b)} shows the SER performance comparison between ADMM and GSD  detectors for the SIMO-overloaded CDMA system. Observe from   Fig.~\ref{Fig. 7(b)} that, at low SNR regions, GSD outperforms the ADMM with a maximum of around 2~dB SNR gain at SER=$10^{-4}$. As the SNR increases, GSD performance degrades compared to ADMM. Unlike GSD, as   $M$ increases, ADMM performance is enhanced by doubling the number of received antennas as shown in Fig.~\ref{Fig. 7(b)}. Note that  ADMM performance improves as the number of observations increases.

\vspace{-5mm}

\subsection{SER performance: Imperfect channel state information (CSI) }
All the above simulations are analyzed by considering perfect CSI at the receiver. 
In practical scenarios, obtaining perfect CSI at the receiver is not feasible due to channel estimation errors (CEEs).
 Thus, the proposed detector's performance in the presence of CEEs is an important metric to show its practical feasibility. An imperfect channel can be modeled as \cite{Wang_Cheng} \begin{align}\hat{\mathbf{H}}=\mathbf{H}+e \ \mathbf{\Omega} \label{eq:CEE}\end{align}
where $\mathbf{\Omega}$ is the CEE and is considered to be uncorrelated with $\mathbf{H}$.  The entries of $\mathbf{\Omega}$ are independent and i.i.d. complex Gaussian random variables with zero mean and unit variance, i.e., $\mathcal{CN}(0,1)$. The quantity $e$ in (\ref{eq:CEE}) determines the variance of the CEE. Fig. \ref{Fig.8} depicts the performance of the proposed ADMM-based detector for the imperfect CSI with $e=0~\%, e=5~\%, e=10~\%$, and $e=20~\%$. The simulations for SCMA and DCMA systems are shown in Fig.~\ref{Fig. 8(a)} and Fig.~\ref{Fig. 8(b)}, respectively. Observe that the impact of the CEE on the ADMM-based detector's performance is minimal. Therefore, the ADMM-based detector can be applied in imperfect CSI scenarios.


\subsection{Convergence of ADMM and selection of parameters}
The convergence analysis of the ADMM-based detector in the DCMA system is performed based on the Monte Carlo simulations. 
Fig.~\ref{Fig. 9 (a)} shows the impact of ADMM iterations on the  SER performance. The proposed detector exploits the iterative nature of ADMM to converge.
\pgfplotsset{every semilogy axis/.append style={
line width=0.7 pt, tick style={line width=0.7pt}}, width=8cm,height=7cm, 
legend style={font=\scriptsize},
legend pos= north east}
\begin{figure*}[htb!]
\centering
 \begin{subfigure} [b]{0.42\textwidth}
        \centering
        \resizebox{\linewidth}{!}{  
\begin{tikzpicture}[new spy style]
\begin{semilogyaxis}[xmin=2, xmax=25, ymin=1e-07, ymax=1e-04,
xlabel={Number of iterations ($T$)},
ylabel={SER},
grid=both,
grid style={dotted},
legend cell align=left,
legend entries={ {$\frac{E_{\mathrm b}}{N_0}$=15dB},{$\frac{E_{\mathrm b}}{N_0}$=18dB},{$\frac{E_{\mathrm b}}{N_0}$=20dB}},
cycle list name=black white
]
\addplot [red,mark=diamond , mark size=3 pt]table [x={T}, y={15dB}]
{data/ADMM_conv.txt};
\addplot [black,mark=square , mark size=3 pt]table [x={T}, y={18dB}]
{data/ADMM_conv.txt};
\addplot [blue,mark=o , mark size=3 pt]table [x={T}, y={20dB}]
{data/ADMM_conv.txt};

\end{semilogyaxis}
\end{tikzpicture}
}
\caption{ \footnotesize{Iterations  vs. SER with $\lambda=150~\%, M=4, N_{\mathrm r}=4$} }
\label{Fig. 9 (a)}
\end{subfigure}
\pgfplotsset{every semilogy axis/.append style={
line width=0.7 pt, tick style={line width=0.7pt}}, width=8cm,height=7cm, 
legend style={font=\scriptsize},
legend pos= north east}
\begin{subfigure}[b]{0.42\textwidth}
        \centering
        \resizebox{\linewidth}{!}{
\begin{tikzpicture}[new spy style]
\begin{semilogyaxis}[xmin=0, xmax=300, ymin=1e-08, ymax=1e-04,
xlabel={$\gamma$},
ylabel={SER},
grid=both,
grid style={dotted},
legend cell align=left,
legend entries={ {$\frac{E_{\mathrm b}}{N_0}$=15dB},{$\frac{E_{\mathrm b}}{N_0}$=18dB},{$\frac{E_{\mathrm b}}{N_0}$=20dB}},
cycle list name=black white
]
\addplot [red,mark=diamond , mark size=3 pt]table [x={al}, y={SER1}]
{data/al_VS_SER.txt};
\addplot [black,mark=square , mark size=3 pt]table [x={al}, y={SER3}]
{data/al_VS_SER.txt};
\addplot [blue,mark=o , mark size=3 pt]table [x={al}, y={SER2}]
{data/al_VS_SER.txt};

\end{semilogyaxis}
\end{tikzpicture}
}

\caption{ \footnotesize{ $\gamma$ vs. SER with $\lambda=150~\%, M=4, N_{\mathrm r}=4$}  }
\label{Fig. 9 (b)}
\end{subfigure}
\label{Fig. 9}
\caption{ \footnotesize {Impact of $T$ and $\gamma$ on the ADMM-based detector's SER performance. } }
\end{figure*}
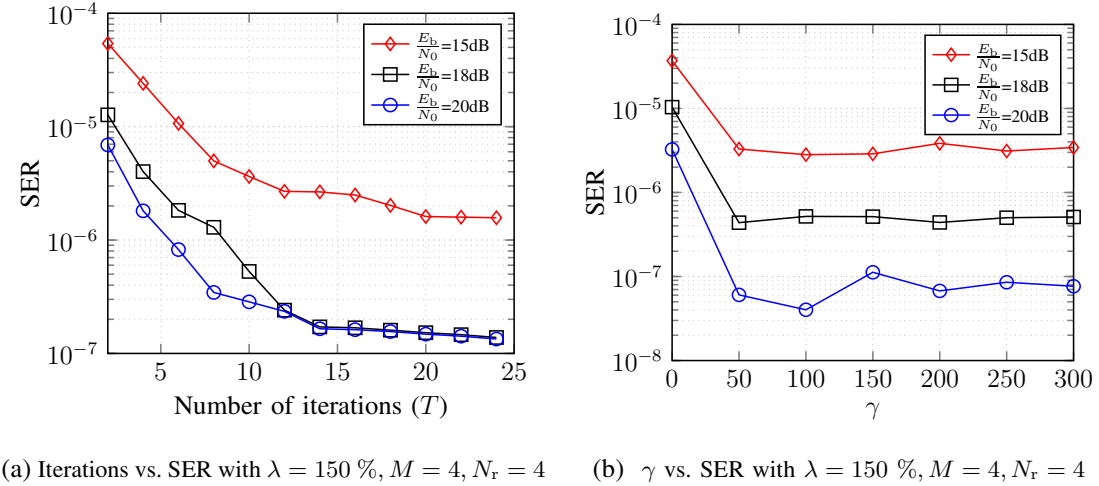
The SER performance improves as the number of iterations ($T$) increases. The improvement in SER becomes marginal after a certain number of iterations. For all considered $\frac{E_{\mathrm b}}{N_0}$ values, ADMM converges after 15 iterations, as shown in Fig.~\ref{Fig. 9 (a)}. 
The penalty parameters $\{\gamma_j\}_{j=1}^J$ and $\rho$ are selected to maximize the SER performance. The augmented Lagrangian parameter $\rho$ is selected as the reciprocal of SNR~\cite{Shahabuddin}. 
The impact of $\gamma$  on SER performance in the ADMM-based detector is analyzed through Monte Carlo simulations, and the results are given in Fig. \ref{Fig. 9 (b)}. For low values of  $\gamma$, the ADMM performance degrades, and for large values, the variation in SER becomes inconsequential. When  $\gamma$ is close to zero, the impact of penalty terms is nullified. Thus, the alternative variables in the penalty function lose importance in minimizing the objective function. Observe from the Fig. \ref{Fig. 9 (b)} that the ADMM performs better in the range $\{\gamma_j\}_{j=1}^J\in [50\hspace{2mm} 100]$.

\section{Conclusions and Future Scopes}\label{sec6}
This paper proposed new system models and  a low-complexity iterative linear detector for large-scale MIMO-CD-NOMA systems. 
The optimal ML detection is converted into a sharing problem, which is efficiently  solved via the ADMM algorithm using distributed optimization framework. 
The proposed ADMM-based detector enables the detection of large-scale MIMO-CD-NOMA systems with high overloading factors ($\lambda$) and modulation orders ($M$) while maintaining low complexity.
By leveraging the proposed ADMM, the CD-NOMA systems achieve significantly increased connectivity.
Further, the impact of ADMM parameters, such as the number of iterations and penalty parameters, is analyzed.
Exhaustive simulation results are presented to validate the effectiveness of the proposed ADMM-based detector when compared to conventional detectors such as MPA, GSD, and MMSE detectors. In addition, the results demonstrate that the ADMM-based detector offers excellent performance with low complexity across various CD-NOMA system variants.
This paper has explored the concept of ADMM detection specifically for uncoded CD-NOMA systems.
Designing a soft decision detector using the ADMM algorithm for coded CD-NOMA systems is an interesting future work.



\bibliographystyle{IEEEtran}
\bibliography{Bibfile,bibJournalList}

\begin{thebibliography}{10}
\providecommand{\url}[1]{#1}
\csname url@samestyle\endcsname
\providecommand{\newblock}{\relax}
\providecommand{\bibinfo}[2]{#2}
\providecommand{\BIBentrySTDinterwordspacing}{\spaceskip=0pt\relax}
\providecommand{\BIBentryALTinterwordstretchfactor}{4}
\providecommand{\BIBentryALTinterwordspacing}{\spaceskip=\fontdimen2\font plus
\BIBentryALTinterwordstretchfactor\fontdimen3\font minus
  \fontdimen4\font\relax}
\providecommand{\BIBforeignlanguage}[2]{{%
\expandafter\ifx\csname l@#1\endcsname\relax
\typeout{** WARNING: IEEEtran.bst: No hyphenation pattern has been}%
\typeout{** loaded for the language `#1'. Using the pattern for}%
\typeout{** the default language instead.}%
\else
\language=\csname l@#1\endcsname
\fi
#2}}
\providecommand{\BIBdecl}{\relax}
\BIBdecl

\bibitem{taherzadeh2014scma}
M.~Taherzadeh, H.~Nikopour, A.~Bayesteh, and H.~Baligh, ``{SCMA} codebook
  design,'' in \emph{Proc, IEEE 80th Veh. Technol. Conf. (VTC-Fall)}, 2014, pp.
  1--5.

\bibitem{Liu}
Y.~L.-L. Liu, Zilong, ``Sparse or {D}ense: A comparative study of code-domain
  {NOMA} systems,'' \emph{IEEE Trans.\ Wireless Commun.}, vol.~20, no.~8, pp.
  4768--4780, 2021.

\bibitem{chockalingam}
A.~Chockalingam and B.~S. Rajan, \emph{Large MIMO Systems}.\hskip 1em plus
  0.5em minus 0.4em\relax Cambridge University Press, 2014.

\bibitem{Lim}
S.-C. Lim, N.~Kim, and H.~Park, ``Uplink {SCMA} system with multiple
  antennas,'' \emph{IEEE Trans.\ Veh.\ Technol.}, vol.~66, no.~8, pp.
  6982--6992, 2017.

\bibitem{Elkawafi}
S.~Elkawafi, A.~Younis, and R.~Mesleh, ``Performance analysis of sparse code
  multiple access {MIMO} systems,'' in \emph{2019 IEEE 30th Int.\ Symp.\ Per.\
  Indoor Mobile Radio Commun (PIMRC)}, 2019, pp. 1--6.

\bibitem{Pan}
Z.~Pan, J.~Luo, J.~Lei, L.~Wen, and C.~Tang, ``Uplink spatial modulation {SCMA}
  system,'' \emph{IEEE Commun.\ Lett.}, vol.~23, no.~1, pp. 184--187, 2019.

\bibitem{Zhang}
C.~Zhang, Y.~Luo, and Y.~Chen, ``A {L}ow-complexity {SCMA} {D}etector based on
  {D}iscretization,'' \emph{IEEE Trans.\ Wireless Commun.}, vol.~17, no.~4, pp.
  2333--2345, 2018.

\bibitem{Yang_Lin}
L.~Yang, Y.~Liu, and Y.~Siu, ``Low complexity message passing algorithm for
  {SCMA} system,'' \emph{IEEE Commun.\ Lett.}, vol.~20, no.~12, pp. 2466--2469,
  2016.

\bibitem{Dai}
J.~Dai, G.~Chen, K.~Niu, and J.~Lin, ``Partially active message passing
  receiver for {MIMO-SCMA} systems,'' \emph{IEEE Wireless Commun. \ Lett.},
  vol.~7, no.~2, pp. 222--225, 2018.

\bibitem{Du}
Y.~Du, B.~Dong, Z.~Chen, P.~Gao, and J.~Fang, ``Joint sparse graph-detector
  design for downlink {MIMO-SCMA} systems,'' \emph{IEEE Wireless Commun. \
  Lett.}, vol.~6, no.~1, pp. 14--17, 2017.

\bibitem{KD_MIMO}
S.~Sharma, K.~Deka, and B.~B. Lozano, ``{Low-complexity Detection for Uplink
  Massive MIMO SCMA Systems},'' \emph{IET Communications}, vol.~15, no.~1, pp.
  51--59, 2021.

\bibitem{Wei}
F.~Wei and W.~Chen, ``Low complexity iterative receiver design for sparse code
  multiple access,'' \emph{IEEE Trans.\ Commun.}, vol.~65, no.~2, pp. 621--634,
  2017.

\bibitem{Yang}
S.~Yang and L.~Hanzo, ``Fifty years of {MIMO} detection: The road to
  large-scale {MIMO}s,'' \emph{IEEE Commun.\ Surveys Tuts.}, vol.~17, no.~4,
  pp. 1941--1988, 2015.

\bibitem{boyd}
S.~Boyd, N.~Parikh, E.~Chu, B.~Peleato, and J.~Eckstein, \emph{Distributed
  Optimization and Statistical Learning via the Alternating Direction Method of
  Multipliers}.\hskip 1em plus 0.5em minus 0.4em\relax Now Foundations and
  Trends, 2011.

\bibitem{Zhouchen}
C.~F. Zhouchen~Lin, Huan~Li, \emph{Alternating Direction Method of Multipliers
  for Machine Learning}.\hskip 1em plus 0.5em minus 0.4em\relax Springer, 2022.

\bibitem{Barman}
S.~Barman, X.~Liu, S.~C. Draper, and B.~Recht, ``Decomposition methods for
  large scale {LP} decoding,'' \emph{IEEE Trans.\ Inf.\ Theory}, vol.~59,
  no.~12, pp. 7870--7886, 2013.

\bibitem{Banihashemi}
H.~Wei and A.~H. Banihashemi, ``{ADMM} check node penalized decoders for {LDPC}
  codes,'' \emph{IEEE Trans.\ Commun.}, vol.~69, no.~6, pp. 3528--3540, 2021.

\bibitem{Cirik}
A.~C. Cirik, N.~Mysore~Balasubramanya, and L.~Lampe, ``Multi-user detection
  using {ADMM}-based compressive sensing for uplink grant-free {NOMA},''
  \emph{IEEE Wireless Commun. \ Lett.}, vol.~7, no.~1, pp. 46--49, 2018.

\bibitem{Shahabuddin}
S.~Shahabuddin, I.~Hautala, M.~Juntti, and C.~Studer, ``{ADMM}-based
  infinity-norm detection for massive {MIMO}: Algorithm and {VLSI}
  architecture,'' \emph{IEEE IEEE Trans.\ Very Large Scale Integr.\ (VLSI)
  Syst.}, vol.~29, no.~4, pp. 747--759, 2021.

\bibitem{Zhang_Quan}
Q.~Zhang, J.~Wang, and Y.~Wang, ``Efficient {QAM} signal detector for {M}assive
  {MIMO} systems via {PS/DPS-ADMM} approaches,'' \emph{IEEE Trans.\ Wireless
  Commun.}, vol.~21, no.~10, pp. 8859--8871, 2022.

\bibitem{Wing-Kin}
W.-K. Ma, T.~Davidson, K.~M. Wong, Z.-Q. Luo, and P.-C. Ching,
  ``Quasi-maximum-likelihood multiuser detection using semi-definite relaxation
  with application to synchronous {CDMA},'' \emph{IEEE Trans.\ Signal
  Process.}, vol.~50, no.~4, pp. 912--922, 2002.

\bibitem{Hoshyar}
R.~Hoshyar, F.~P. Wathan, and R.~Tafazolli, ``Novel low-density signature for
  synchronous {CDMA} systems over {AWGN} channel,'' \emph{IEEE Trans.\ Signal
  Process.}, vol.~56, no.~4, pp. 1616--1626, 2008.

\bibitem{Nikopour}
H.~Nikopour and H.~Baligh, ``Sparse code multiple access,'' in \emph{2013 IEEE
  24th Annual Int.\ Symp.\ Per.\ Indoor Mobile Radio Commun (PIMRC)}, 2013, pp.
  332--336.

\bibitem{Boutros}
J.~Boutros and E.~Viterbo, ``Signal space diversity: a power- and
  bandwidth-efficient diversity technique for the rayleigh fading channel,''
  \emph{IEEE Trans.\ Inf.\ Theory}, vol.~44, no.~4, pp. 1453--1467, 1998.

\bibitem{marzetta}
T.~L. Marzetta and H.~Yang, \emph{Fundamentals of massive {MIMO}}.\hskip 1em
  plus 0.5em minus 0.4em\relax Cambridge University Press, 2016.

\bibitem{Mesleh}
R.~Y. Mesleh, H.~Haas, S.~Sinanovic, C.~W. Ahn, and S.~Yun, ``Spatial
  modulation,'' \emph{IEEE Trans.\ Veh.\ Technol.}, vol.~57, no.~4, pp.
  2228--2241, 2008.

\bibitem{Yusha}
Y.~Liu, L.-L. Yang, and L.~Hanzo, ``Spatial modulation aided sparse
  code-division multiple access,'' \emph{IEEE Trans.\ Wireless Commun.},
  vol.~17, no.~3, pp. 1474--1487, 2018.

\bibitem{Yaoyue}
H.~Yaoyue, P.~Zhiwen, L.~Nan, and Y.~Xiaohu, ``Multidimensional constellation
  design for spatial modulated {SCMA} systems,'' \emph{IEEE Trans.\ Veh.\
  Technol.}, vol.~70, no.~9, pp. 8795--8810, 2021.

\bibitem{Studer}
C.~Studer, A.~Burg, and H.~Bolcskei, ``Soft-output sphere decoding: algorithms
  and {VLSI} implementation,'' \emph{IEEE {J.} Sel.\ Areas Commun.}, vol.~26,
  no.~2, pp. 290--300, 2008.

\bibitem{Cui}
T.~Cui and C.~Tellambura, ``An efficient generalized sphere decoder for
  rank-deficient {MIMO} systems,'' \emph{IEEE Commun.\ Lett.}, vol.~9, no.~5,
  pp. 423--425, 2005.

\bibitem{Hassibi}
B.~Hassibi and H.~Vikalo, ``On the sphere-decoding algorithm {I}. {E}xpected
  complexity,'' \emph{IEEE Trans.\ Signal Process.}, vol.~53, no.~8, pp.
  2806--2818, 2005.

\bibitem{Vikas}
V.~Vikas, A.~Rajesh, K.~Deka, and S.~Sharma, ``A comprehensive technique to
  design {SCMA} codebooks,'' \emph{IEEE Commun.\ Lett.}, vol.~26, no.~8, pp.
  1735--1739, 2022.

\bibitem{Wang_Cheng}
C.~Wang, E.~K. Au, R.~D. Murch, W.~H. Mow, R.~S. Cheng, and V.~Lau, ``On the
  performance of the {MIMO} zero-forcing receiver in the presence of channel
  estimation error,'' \emph{IEEE Trans.\ Wireless Commun.}, vol.~6, no.~3, pp.
  805--810, 2007.

\end{thebibliography}

\end{document}